\documentclass[useAMS]{mn2e}
\input epsf
\usepackage{lscape}
\usepackage{rotating}

\def\two{\,{\sc ii}}

\input psfig.sty
\title[Mid Infra Red Imaging]{MSX Mid Infrared Imaging of Massive Star Birth 
Environments. I: Ultracompact H\,{\bf\sc ii} Regions}
\author[Paul A. Crowther and Peter S. Conti] {Paul A. 
Crowther\thanks{E-mail:  pac@star.ucl.ac.uk (PAC) 
pconti@jila.colorado.edu (PSC)}, and Peter S. Conti $^\star$\\
Department of Physics and
Astronomy, University College London, Gower Street, London, WC1E 6BT, UK; \\
JILA and APS Department, University of Colorado, Boulder CO~80303 USA}
\date{Accepted. Received; in original form}

\pagerange{\pageref{firstpage}--\pageref{lastpage}}
\pubyear{2002}
\begin{document}

\maketitle

\label{firstpage}

\begin{abstract}
We present mid-IR 21$\mu$m images of a sample of radio selected
Ultracompact H\two\ (UCH\,{\sc ii})  
regions, obtained with the Midcourse Space Experiment (MSX). All,
with one possible exception, are detected at mid-IR wavelengths, 
sampling the warm dust emission of the cocoons of the OB star central
exciting sources.  Many of the UCH\,{\sc ii} regions have nearby (up to $\approx$
few pc distant) companion dust emission sources, which represent other
potential star birth sites.  In some objects the companion dominates the
IRAS point source catalogue entry for the UCH\,{\sc ii} region.
We compare the
mid- and far-IR dust emission, measuring the embedded hot star luminosity,
with published UCH\,{\sc ii} radio emission, measuring the Lyman continuum 
luminosity.  
We find a spectral type dependence, as predicted by the 
standard model of an ultracompact ionized hydrogen region, surrounded by a natal
dust shell, with some scatter, which
can be understood by consideration of: 1) dust absorption of some fraction
of the emitted Lyman continuum photons; 2) fainter companion stars within 
the UCH\,{\sc ii} region;  3) the structure of the UCH\,{\sc ii} regions differing from 
star to star. Overall, the higher spatial resolution offered by MSX 
alleviates difficulties often encountered by comparison of IRAS far-IR 
fluxes with radio derived ionizing fluxes for UCH\,{\sc ii} regions.
\end{abstract}

\section{Introduction}

Massive O and B-type stars are born individually in molecular clouds or
collectively within giant molecular clouds (GMCs), which contain
substantial amounts of dust.  From the collapse and fractionation of the
GMC to the final product of a visible star (or a cluster) a number of
physical processes are occurring. These include, but are not limited to,
the following stages for the {\em star}: rapid accretion, the (possible)
formation of a disk, slow accretion, the initiation of the zero age main
sequence (ZAMS), and normal main
sequence evolution.  At the same time, but {\em not} with identical
timescales, its surrounding {\em environment} is evolving: the molecular
hydrogen is dissociated then ionized into a gradually expanding H\two\
region; the ice mantles of the dust are evaporated, and the natal cocoon
expands, lowering its overall density and opacity. For a recent summary
of this general topic, 
the reader is referred to {\it The Earliest Stages of Massive Star Birth} (Crowther 2002). 

The wavelength at which
the material becomes optically thin to the stellar radiation shifts to
shorter and shorter values from the IR. The dust in the cocoon and in the
surroundings absorbs and reprocesses the stellar radiation and re-emits it
in the far and mid IR (FIR--MIR) regions (from ~1 mm down to 10$\mu$m).
As this evolution proceeds, we would expect to {\em observe} the following
phenomena for each individual stellar object:

$\bullet$ A `Hot Core' phase (e.g., Kurtz et al. 2000) in which a source
of IR dust emission surrounding a buried star is found. The star is
sufficiently luminous to heat the dust but it is not (yet?) hot enough to
produce sufficient Lyman Continuum photons. Alternatively, 
the infalling material is ``quenching'' the formation of an
H\two\ region (e.g. Osorio et al. 1999).

$\bullet$ The star forms an ultracompact H\two\ (UCH\,{\sc ii}) region 
(e.g.,
Churchwell 1999ab) which can be identified from its free-free radio
emission. The ionized hydrogen is surrounded by a natal dust cocoon which
radiates in the FIR--MIR.

$\bullet$ The UCH\,{\sc ii} region expands, becoming a compact H\two\ 
(CH\two) 
region, as the dust evaporates and the overlying cocoon expands and 
thins out (as in, e.g., W49A -- Conti \& Blum 2002). The FIR--MIR emission 
gradually diminishes.

$\bullet$ The dust becomes optically thin at visible wavelengths.
Eventually the star no longer has any surrounding natal material (e.g., a 
disk). We will identify this time as the end of the birth process for an 
individual massive star.

Two key parameters of the birth processes for massive stars are those of
the H\two\ region and those of the dust emission.  The former can be 
estimated from
the Lyman continuum luminosity, or extreme ultraviolet (EUV) radiation, which is highly
dependent on the stellar temperature (e.g. Smith, Norris \& Crowther 
2002); the latter  from the overall stellar
luminosity (primarily the UV radiation) along with the dust properties
(thickness, extent, and distance from the star). As the birth process goes
forward, we would expect both of these parameters to change in response to
the stellar evolution and the varying dust in the surrounding cocoon. The
dust emission will undergo the largest change as it responds to the
evolution of the environment. The Lyman continuum and the UV luminosity
should not change dramatically as once the star ``turns on'' (at the ZAMS)
as one would expect that substantial accretion will come to a halt (thus
$T_{\rm eff}$ and $L_{\rm bol}~ \approx$  constant). For cases in which 
the central
O star is obscured by hundreds of magnitudes of visual extinction, the situation
is much simpler at mid-IR wavelengths, where the extinction is several hundred
times lower.

Until recently, there had not been any MIR Galactic plane surveys of
medium spatial resolution\footnote{IRAS had $30'', 30'', 1', 2'$ 
resolutions for the $12\mu$m, $25\mu$m, $60\mu$m, $100\mu$m filters, 
respectively}.  The
Midcourse Space Experiment (MSX) remedied this situation, with a complete
Galactic plane survey within $|b|\leq$5$^{\circ}$ (Price et al. 2001) at
18$''$ spatial resolution. Ultimately, the Space Infrared Telescope Facility
(SIRTF) will achieve a much higher spatial resolution. For the moment,
however,  MSX  permits the comparison  between radio and MIR fluxes of almost
all known massive star birth sites, so as to better
understand the relationship between the H\two\ regions and the dust 
cocoons.  
In this paper we shall begin with a study of a sample of the 
UCH\,{\sc ii} regions
in our Galaxy, which are generally believed to be excited by one or
a few dominant stars,
 and
ought to represent the simplest cases to investigate. A subsequent paper
will consider our results for H\two\ and giant H\two\ (GH\two) regions.

We begin in Section~\ref{obs} by outlining the extraction procedure for
obtaining MIR data from the MSX satellite. In Section~\ref{sam} we list
the sample of UCH\,{\sc ii} objects from the catalogs of Wood \& Churchwell (1989)
and Kurtz, Churchwell \& Wood (1994) that we have studied.  We discuss in
Section~\ref{morph} MIR images at 21 $\mu$m of the UCH\,{\sc ii} regions, and
comment on similarities and differences between the objects. Many objects 
are multiple, with close (within a few pc) companions. In some of
these, the IRAS fluxes of the UCH\,{\sc ii} regions will have been affected by the
nearby objects, in a few cases drastically so.  In Section~\ref{comp} we
first consider the relationships among the MSX medium resolution and IRAS
low resolution data for our sources. After accounting for the multiplicity
of some sources in the IRAS data, and noting a ``colour'' term in the MSX
data, we recover the spectral type dependence of the EUV to UV fluxes of
the sources, as predicted by relatively simple models of UCH\,{\sc ii} regions.  
Conclusions are given in Section~\ref{dis}.

\begin{sidewaystable*}
\begin{tabular}{r@{\hspace{3mm}}l@{\hspace{2mm}}
r@{\hspace{2.5mm}}r@{\hspace{2.5mm}}r@{\hspace{2.5mm}}r@{\hspace{2.5mm}}
r@{\hspace{2.5mm}}l@{\hspace{2.5mm}}r@{\hspace{2.5mm}}c@{\hspace{2.5mm}}r@{\hspace{2.5mm}}r@{\hspace{2.5mm}}
r@{\hspace{2.5mm}}c@{\hspace{2.5mm}}c@{\hspace{2.5mm}}c@{\hspace{2.5mm}}l}                
\noalign{{\bf Table 1:} Mid-IR (MSX/IRAS) and radio (VLA/Arecibo) fluxes (in Jy) for 
Galactic UCH\,{\sc ii} regions from Wood \& Churchwell (1989, WC) and Kurtz et al. 
(1994, KCW). UCH\,{\sc ii} regions with multiple radio peaks are summed together (e.g. G10.47+03 A,B,C).}\\
\hline
UCH\,{\sc ii} G&IRAS& 12$\mu$m & 21$\mu$m&25$\mu$m&100$\mu$m& 2\,cm & 6\,cm      & 
$d$   &Ref& $\log N$&
21$\mu$m/ & 25$\mu$m/ & 21$\mu$m/& 100$\mu$m & Radio&MSX\\
     & & MSX  &MSX  &IRAS&IRAS& VLA&VLA& kpc &    &Ly\,C  & 12$\mu$m$^{a}$  & 21$\mu$m$^{a}$ & 2\,cm$^{a}$
& 2\,cm$^{a}$ & morph  &morph \\
\hline
[WC89] 5.476--0.243 &17559-2420& 23&  89 & 194& 2170&     & 0.12 &14.3&1& 
48.35   &0.59&0.34&--- &--- &Core-Halo&Extended\\
{}[WC89] 5.885--0.392* &17574-2403&128& 776 &2190&26780& 6.54& 2.09 & 
2.0&4& 
48.54   &0.78&0.45&2.08&3.61&Shell&Multiple\\
{}[WC89] 5.972--1.174 &18006-2422&120& 762 &1842& 9036&     & 0.28 & 
2.7&2& 
47.71   &0.80&0.65&--- &--- &Core-Halo&Extended\\
{}[WC89] 8.669--0.356* &18034-2137&$<$0.8:&10& 154& 5221&0.57& 0.63 & 
4.6&2& 
48.09 &$>$1.1:&1.19&1.23&    &Core-Halo&Double\\
{}[WC89] 10.304--0.147*&18060-2005& 53& 189 &1042&12010&     & 0.51 & 
6.0&2& 
48.21   &0.55&0.74&--- &--- &Cometary &Multiple\\
\\
{}[WC89] 10.460+0.032* &18056-1952&5.3&  18 & 106&10160&     & 0.12 & 
5.8&2& 
48.22   &0.53&0.77&--- &---  &Spherical &Double\\
{}[WC89] 10.472+0.027* &18056-1952&3.1&  19 & 106&10160&    & 0.05 & 5.8&2& 
47.20    &0.79&0.75&--- &---  &Spherical &Double\\
{}[WC89] 10.623--0.384*&18075-1956&3.8&  41 & 148&21370&2.13& 1.21 &17.5&2& 
49.82    &1.03&0.56&1.28&     &Core-Halo&Multiple\\
{}[KCW94] 10.841--2.592&18162-2048& 22& 149 & 347& 3709&0.009&      & 
1.9&3& 
45.00   &0.83&0.37&4.23&5.63 &  &Multiple  \\
{}[WC89] 11.938--0.616&18110-1854& 12&  99 & 222& 4930&1.07& 0.88 & 5.2&2& 
48.28    &0.92&0.35&1.96&3.66 &Cometary & Double\\
\\
{}[WC89] 12.209--0.103*&18097-1825A&2.7&   7&  10& 4221&0.23& 0.17 &13.5&2& 
48.62    &0.41&0.16&1.46&    &Cometary &Multiple\\
{}[WC89] 12.429--0.049&18099-1811&1.8:&  3 &  8 &  510&    & 0.04 &16.7&1& 
47.99    &0.2:&0.5:&--- &--- &Cometary &Single\\
{}[KCW94] 18.146--0.284&18222-1317&43 & 138 & 394& 8494&0.011&      & 
4.2&3& 
46.82   &0.51&0.45&4.10&5.89&  &Multiple  \\
{}[WC89] 19.608--0.234&18248-1158&47 &  195& 407& 7093&0.31& 0.80 & 3.5&2& 
47.62    &0.62&0.32&2.80&4.36&Cometary &Extended\\
{}[WC89] 20.080--0.135*&18253-1130&4.6&   21&  76& 2761&0.51& 0.38 & 3.4&2& 
47.78    &0.66&0.56&1.61&3.73&Shell&Double \\
\\
{}[WC89] 23.455--0.201*&18319-0834&$<$1.2:&2:&122& 9595&    & 0.01 & 9.0&1& 
46.84   &0.2::&1.8:&--- &--- &Spherical &Multiple \\
{}[WC89] 23.711+0.171* &18311-0809&20 & 78  & 184& 2904&    & 0.15 & 8.9&2& 
48.02   &0.59 &0.37&--- &--- &Core-Halo&Double \\
{}[WC89] 25.716+0.049*&18353-0628&5.0& 31  & 105& 2435&    & 0.02 & 9.3&2& 
47.08   &0.79 &0.53&--- &--- &Spherical &Multiple \\
{}[KCW94] 28.200--0.049&18403-0417& 21& 70  & 178& 3937&0.77&      & 
9.1&3& 
48.47   &0.53 &0.40&1.96&3.71&  &Single  \\
{}[KCW94] 28.288--0.364&18416-0420& 63&473  & 821& 4358&0.94&      & 
3.3&3& 
47.81   &0.88 &0.24&2.70&3.66&  &Multiple   \\
\\
{}[WC89] 29.956--0.016&18434-0242&208&1022 &1697&11670&2.66& 1.37 & 7.4&2& 
49.17   &0.69 &0.22&2.58&3.64&Cometary &Multiple\\
{}[WC89] 30.535+0.021&18443-0210& 13 & 47 &  88& 1067&0.46 & 0.45 &13.8&1& 
48.96  &0.56 &0.27&2.01&3.37&Cometary &Single\\
{}[WC89] 31.414+0.310 &18449-0115& 2.0&15  &  52& 2815&0.48 & 0.38 & 
7.9&2& 
48.49  &0.87 &0.52&1.51&3.76&Core-Halo&Single \\
{}[KCW94] 32.798+0.190 &18479-0005 & 30 &123 & 298& 5473&0.45 &
                                                                        &13.0&3& 48.87  &0.61 &0.38&2.44 &4.08&  &Single \\
{}[WC89] 33.915+0.110 &18502+0051& 30 &137 & 250& 2429&0.36& 0.50 
                   & 8.2&1& 48.54  &0.66 &0.26&2.44&3.68&Core-Halo&Extended \\
\\
{}[WC89] 34.255+0.145 &18507+0110& 106& 434&1106&32460&4.18 & 1.58 
                                                                        & 4.0&2& 48.83   &0.61&0.41&2.02&3.89&Cometary &Multiple \\
{}[WC89] 35.199--1.743&18592+0108&  88&471 &1073&13960& 2.37& 1.93 
                                                                        & 3.1&2& 48.40   &0.73&0.34&2.30&   &Cometary &Extended\\
{}[WC89] 37.545--0.112&18577+0358& 13 & 48 & 106& 1867&     & 0.23 & 
9.9&2& 
48.30   &0.57&0.34&--- &--- &Core-Halo&Extended\\
{}[KCW94] 37.874--0.399&18593+0408& 42 & 129& 304& 4524&3.07&
                                                                         & 9.2&3& 49.18  &0.49 &0.37&1.62&3.18&  &Single  \\
{}[WC89] 43.889--0.783&19120+0917& 15 & 74 & 145& 1511&0.51 &0.36   & 
4.2&2& 
47.96  &0.69 &0.29&2.16&3.47&Cometary &Extended \\
\hline
\end{tabular}  
{\begin{small}
\newline (1) Wood \& Churchwell (1989); (2) Churchwell et al. (1990); 
(3) Kurtz et al. (1994); (4) Acord et al. (1998)
\newline (a) entries are presented in logarithms for ready comparison with figures.
\end{small}}
\end{sidewaystable*}

\setcounter{table}{1}

\begin{sidewaystable*}
\begin{tabular}{r@{\hspace{3mm}}l@{\hspace{2mm}}
r@{\hspace{2.5mm}}r@{\hspace{2.5mm}}r@{\hspace{2.5mm}}r@{\hspace{2.5mm}}
r@{\hspace{2.5mm}}l@{\hspace{2.5mm}}r@{\hspace{2.5mm}}c@{\hspace{2.5mm}}r@{\hspace{2.5mm}}r@{\hspace{2.5mm}}
r@{\hspace{2.5mm}}c@{\hspace{2.5mm}}c@{\hspace{2.5mm}}c@{\hspace{2.5mm}}l}                
\noalign{{\bf Table 1: (continued)}}\\
\hline
UCH\,{\sc ii} G&IRAS& 12$\mu$m & 21$\mu$m&25$\mu$m&100$\mu$m& 2\,cm & 6\,cm      & 
$d$   &Ref& $\log N$&
21$\mu$m/ & 25$\mu$m/ & 21$\mu$m/& 100$\mu$m & Radio&MSX\\
     & & MSX  &MSX  &IRAS&IRAS& VLA&VLA& kpc &    &Ly\,C  & 12$\mu$m$^{a}$  & 21$\mu$m$^{a}$ & 2\,cm$^{a}$
& 2\,cm$^{a}$ & morph  &morph \\
\hline
{}[WC89] 45.071+0.132 &19110+1045& 47 & 234& 494& 7497&0.59& 0.14 
                                                                         & 6.0&2& 48.33  &0.70&0.32&2.60&4.10&Spherical &Single \\
{}[WC89] 45.122+0.132 &19111+1048 & 260&1007&1395& 7497&3.68& 1.17 
                                                                         & 6.9&2& 49.25  &0.59&0.14&2.44&3.31&Cometary &Multiple\\
{}[WC89]
 45.456+0.060 &19120+1103& 71 & 367& 640& 7890&     & 0.42 
                                                                         & 6.6&2& 48.21  &0.71&0.24&---&--- &Cometary &Multiple\\
{}[WC89] 45.466+0.046* &19120+1103& 0.6 & 17 & 640& 7890&    & 0.08 
                                                                         & 6.0&2&        &1.46&1.6:&---&----&  &Multiple\\
{}[KCW94] 48.606+0.024* &19181+1349&9.3 & 33 & 175& 5227&0.06&
                                                                         & 9.7&3& 47.59  &0.55 &0.72&2.74&   &  &Multiple  \\
\\
{}[WC89] 49.490--0.370*&19213+1424& 402&1658&4344&26760&5.35 & 0.50 & 
6.6&2& 
49.37   &0.62 &0.42&2.49&   &Cometary &Multiple\\
{}[WC89] 54.094--0.060&19294+1836& 6.7&22  &  50& 2136&     & 0.002& 
7.9&1& 
46.00   &0.52 &0.36&--- &---&Spherical &Double \\
{}[KCW94] 60.884--0.128&19442+2427& 30 &208 & 425& 5174&0.029&
                                                                        & 2.3&3& 46.50   &0.84 &0.31&3.86&5.26& &Double   \\
{}[WC89] 61.473+0.093 &19446+2505& 64& 458 &1185&13210&    & 0.25 & 6.5&2& 
47.22    &0.85 &0.41&--- &---& Spherical &Extended\\
{}[KCW94] 70.293+1.600* &19598+3324&290& 927 &1780&12980&5.15 &
                                                                        & 8.6&3& 49.29   &0.51 &0.28&2.26&3.38&  &Multiple \\
\\
{}[KCW94] 70.330+1.586* &19598+3324&15& 58  & 1780&12980&0.23 &
                                                                        & 8.0&3& 48.15   &0.60 &1.49&2.41&   &  &Multiple  \\ 
{}[WC89] 75.783+0.343* &20198+3716& 8.2& 45 & 480& 6985&     & 0.04 & 
4.1&2& 
46.78   &0.74 &1.03&--- &---&Cometary &Double\\
{}[WC89] 75.835+0.400 &20197+3722& 109& 656&1225& 6985&     & 0.27 & 
5.5&1& 
47.86   &0.78 &0.27&--- &---&Core-Halo&Double \\ 
{}[KCW94] 76.383--0.621&20255+3712& 164&1050&2510&13130&0.017&      & 
1.0&3& 
45.06   &0.81 &0.38&4.80&5.90 & Spherical&Extended\\
{}[KCW94] 78.438+2.659 &20178+4046&26  & 339& 551& 2877&0.036&      & 
3.3&3& 
46.83   &1.12 &0.21&3.98&4.91&   &Single   \\
\\
{}[KCW94] 81.679+0.537 &...        &27  & 307&    &     &0.88 &      & 
2.0&3& 47.52   &1.06 & ---&2.54& ---& Core-Halo&Double \\
{}[KCW94] 81.683+0.541 &...        &33  & 397&    &     &0.61 &       
&2.0&3& 47.35   &1.08 & ---&2.81& ---&Core-Halo&Double  \\
{}[KCW94] 109.871+2.113&20543+6145& 5.3& 249& 820&20470&0.027&      & 
0.7&3&44.62    &1.67 &0.52&3.97&5.88&Multiple&Double  \\
{}[KCW94] 111.282--0.663&23138+5945&16& 92 & 233& 2164&0.080&      & 
2.5&3&46.68    &0.75 &0.40&3.06&4.43&Core-Halo&Multiple  \\
{}[KCW94]
 111.612+0.374&23133+6050&  38&371 & 581& 2694&0.68 &      & 
5.2&3&48.21    &0.98 &0.19&2.74&3.59&Shell&Single   \\
{}[KCW94] 133.947+1.064&02232+6137& 32 &250 & 536&10600&2.53 &      & 
3.0&3&47.92    &0.89 &0.33&2.00&3.62& Spherical& Single  
\\
{}[KCW94] 139.909+0.197&03035+5819& 25 &237 & 396& 1297&0.018&      & 
4.2&3&46.43    & 0.98&0.22&4.11&4.85& Spherical& Single  \\
{}[KCW94] 192.584--0.041*&06099+1800&$<$1.5:&16& 371& 5285&0.026& & 
2.5&3&46.03   &$>$1.05&1.37&2.79&    &  &Multiple   \\
\hline
\end{tabular}  
{\begin{small}
\newline (1) Wood \& Churchwell (1989); (2) Churchwell et al. (1990); 
(3) Kurtz et al. (1994); (4) Acord et al. (1998)
\newline (a) entries are presented in logarithms for ready comparison with figures.
\end{small}}
\end{sidewaystable*}

\section{MSX Observations}\label{obs}

The U.S. Department of Defense MSX satellite surveyed the entire Galactic
plane in four mid-IR spectral bands, named A, C, D, and E, between 6 and
25$\mu$m at a spatial resolution of $\sim$18.3$''$.  A description of the
telescope and primary instrument, SPIRIT III is given by Price et al.
(2001). The entire area within $\pm$4.5$^{\circ}$ of the Galactic plane
was surveyed at least twice, with four-fold coverage obtained in the
first and fourth quadrants, and to $\pm$3$^{\circ}$ for the remanding
quadrants. The redundancy was sufficient to permit combining the datasets
onto a uniformly spaced grid, such that the inherent spatial resolution
of the instrument was preserved. A total of 1680 95.1$'\times95.1'$
images were created in each of the four mid-IR spectral bands, spaced
90$'$ apart in each coordinate, providing 10$'$ overlap on adjacent
images.  Each image provides radiances on a grid in Galactic latitude and
longitude, with a grid spacing of 6$''$. The image units are of in-band
radiance (W m$^{-2}$ sr$^{-1}$).

\begin{figure} 
\epsfxsize=8.8cm\epsfbox[0 115 504 725]{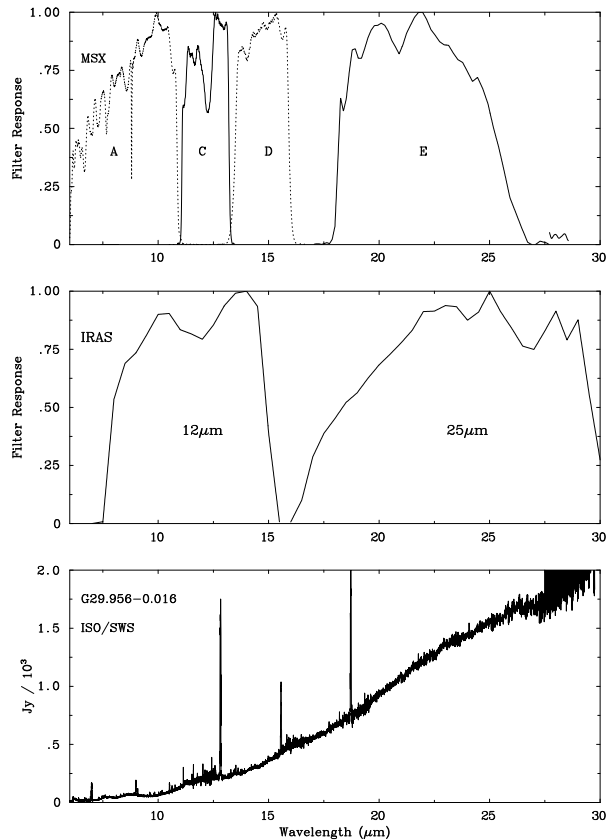}
\caption{Comparison between the broad band, mid-IR filter profiles of MSX,
of which Band C (12$\mu$m) and Band E (21$\mu$m) are used here, and the
IRAS 12 and 25$\mu$m filters, together with the ISO/SWS 
spectrum of the UCH\,{\sc ii} region G29.956--0.016 (Morisset et al. 
2002) for comparison. Strong fine-structure nebular lines include
[Ne\,{\sc ii}] 12.8$\mu$m, [Ne\,{\sc iii}] 15.5$\mu$m and [S\,{\sc iii}]
18.7$\mu$m.}\label{fig0} \end{figure} 

Of the available bands, Band~A is the most sensitive, but its filter
footprint of 6.8--10.8$\mu$m includes the astrophysically strong, broad
silicate band at 9.7$\mu$m and the 8.6$\mu$m PAH feature. 
We therefore selected the longer wavelength
Band~C and Band~E observations. These have isophotal central wavelengths
of $\lambda_c$=12.13$\mu$m and 21.34$\mu$m, respectively, with bandwidth
of 1.72$\mu$m and 6.24$\mu$m.  MSX Bands~C and E are narrower, higher
spatial resolution analogues of the well known IRAS 12$\mu$m and 25$\mu$m
bands as illustrated in Fig.~\ref{fig0}, where ISO/SWS spectroscopy
of the prototypical UCH\,{\sc ii} region G29.956--0.016 is also presented 
(Morisset et al. 2002).  
The calibration and photometric accuracy of MSX are discussed in
detail by Egan et al. (1999). The zero magnitude flux is based on the
Kurucz model for Vega (Cohen et al. 1992), and correspond to 9.259$\times
10^{-13}$ W m$^{-2}$ (Band~C) and 3.555$\times 10^{-13}$ W m$^{-2}$ (Band
E). 

Individual UCH\,{\sc ii} images were obtained from the MSX Image Server at IPAC
{\tt http://irsa.ipac.caltech.edu/applications/MSX/}. For our purposes, we
sought integrated MIR fluxes in Janskys for comparison with radio fluxes.  
Consequently, the MSX point source catalogue was not appropriate, given
the need to use apertures precisely centred on the radio coordinates.
Spatial integration was measured in {\sc gaia} (Draper, Gray \& Berry 
2001)  
using a circular aperture of radius 3 pixel (18$''$). Aperture centres
were selected to precisely mimic those of the corresponding radio
observations. Thin annuli, with 3 pixel radii (18$''$), were used to
correct for the background flux levels, taking care to avoid sources in
the annuli. Spatial integration over the images provides fluxes in units
of W m$^{-2}$, after correction for the 6$''\times 6''$ pixel-area, i.e.
8.4615$\times 10^{-10}$ sr. Division by the filter bandwidth then provides
fluxes in W m$^{-2}$ Hz$^{-1}$ (= 10$^{26}$ Jy). Finally, a multiplication
by 1.113 is required to convert the square area pixels into the correct
Gaussian area (Cohen, priv. comm.).  The total scale factor for Band~C
corresponds to a multiplicative factor of 27.45 in order to convert the
IPAC integrated fluxes to mJy, whilst the corresponding factor is 23.72
for Band~E.

\section{Selection of Sample}\label{sam}

We included all the UCH\,{\sc ii} regions from Wood \& Churchwell (1989) and Kurtz
et al. (1994) for which 2 or 6\,cm radio fluxes (and so ionizing
fluxes) exist, and contain MSX datasets for which data could be extracted.  
We have omitted G15.042--0.676 since its MIR flux is totally dominated by
the surrounding GH\two\ region M17. Distances have been
kinematically determined from Galactic rotation model by
Churchwell, Walmsley \& Cesaroni (1990) or Kurtz et al. (1994), except 
where noted. Table
1 lists the parameters of the UCH\,{\sc ii} regions we have studied in the MIR.
The various columns will be discussed in due course.

\section{Mid-IR morphologies}\label{morph}

\subsection{Discussion of individual objects}

Figures A.1--~8 in the Appendix contain 21$\mu$m images 
of our sources 
at spatial scales of 10$' \times 10'$, having first been 
transformed from (l, b) to (RA, Dec.) for epoch J2000.0.
We have indicated the physical scales, using the distances
derived from Churchwell et al. (1990) or Kurtz et al. (1994).  

We have
also made corresponding images of the 12$\mu$m data from MSX (not
presented here).  For the
most part, these appear similar with the exception that 1) several UCH\,{\sc ii}
sources are not visible at this wavelength and 2) some other point sources
are seen at the shorter wavelength but not the longer one. The former are
inordinately red sources with a cool dust origin, whilst the latter are
probably stars, which would typically be brighter at 12$\mu$m. NIR images
of the central 0.5$' \times 0.5'$ for many UCH\,{\sc ii} regions are presented
by Hanson, Luhman \& Rieke (2002).

To facilitate the presentations below, we have indicated with an asterisk
(*) those sources for which the lower resolution IRAS photometry and its
Point Source Catalog (PSC) will be affected by a brighter nearby companion
as suggested by our images.

\subsubsection{[WC89] G5.476--0.243}

The 21$\mu$m image of G5.476 is presented in Fig.~A1(a), revealing a
single core, but within a spatially extended source.
The nearest mid-IR source is much fainter, and is probably unrelated as it lies
150$''$ (at least $\sim$10 pc) away to the north east.

\subsubsection{[WC89] G5.885--0.392*}

G5.885, alias W28 A2(1), 
is amongst the brightest UCH\,{\sc ii} of our sample at mid-IR wavelengths.
The mid-IR morphology is shown in Fig.~A1(b), and reveals an elongation to the SE,
probably due to another source. From MSX, a separate
spatially extended region is only 2.5$'$ away ($\sim$ 1.5\,pc) which
also contains a bright point source at G5.898--0.441 (same IRAS source as
G5.885).  Kim \& Koo (2001) have recently presented 21cm VLA observations
with a spatial resolution comparable to MSX, revealing a morphology
very similar to Fig.~A1, with G5.885 corresponding to their more compact
western source. Since the larger eastern source is a radio emitter it
is not a hot core.
These are embedded within a large region (14 $\times$ 9$'$) of weak emission.

Note that we have adopted 2\,kpc for the distance to G5.89 as
determined by Acord, Churchwell \& Wood (1998). This object was used as 
the basis of
the standard model of an UCH\,{\sc ii} region by Wolfire \& Churchwell (1994).
 Feldt et al. (1999) discuss arcsecond resolution NIR and MIR images of G5.885.

\subsubsection{[WC89] G5.972--0.174} 

G5.972, within the Lagoon Nebula (M8), 
appears as an extended mid-IR source in the MSX image
Fig.~A1(c), although it is uniquely surrounded by a very large,
extended halo, several parsec in extent.
The flux enclosed within a radius of 36 arcsec is a factor
of 2 times larger than the nominal (18 arcsec radius) value.
 Stecklum et al. (1998) discuss high spatial resolution optical, 
IR and radio observations of G5.972. 21cm data,
showing the larger scale structure, are presented by Kim \& Koo (2001), revealing 
a very large (14$'$ $\times$ 10$'$) extended envelope surrounding the UCH\,{\sc ii} region, 
including an arclike structure originating from the SE, that has no obvious mid-IR 
counterpart.

\subsubsection{[WC89] G8.669--0.356*}

From our sample, G8.669 has amongst the reddest 12--21$\mu$m index, such
that it is barely detectable at 12$\mu$m, with a corresponding low dust
temperature. The mid-IR flux measured by IRAS at the location of
G8.669--0.356 is actually dominated by a probable hot core some 40$''$ or
0.9 pc away - Fig.~A1(d).  Consequently, previously derived
luminosities obtained from the IRAS PSC for this UCH\,{\sc ii} represent a strong
overestimate, i.e. 19\,Jy was measured by IRAS at 12$\mu$m versus $<$0.8Jy
obtained here for the UCH\,{\sc ii} region.

\subsubsection{[WC89] G10.304--0.147*}

The mid-IR image of G10.304, within the W31 H\two\ complex, is shown in 
Fig.~A1(e), and
reveals it to be spatially extended, with a faint dust tail extending north-west, 
with a probably related strong extended source, centred 70$''$ away ($\sim$2 pc) 
to the NE  at G10.321--0.157. IRAS\,18060--2005 contains both IR sources, such that
the IR luminosity of G10.304 has previously been strongly overestimated.
Kim \& Koo (2001) present 21cm VLA observations of G10.304, also revealing strong,
concentrated emission in the bright mid-IR source to the NE.  
A faint radio envelope surrounds both components, extending 13$'$ $\times$ 5$'$ to the NW and SE.

\subsubsection{[WC89] G10.472+0.032* and 10.460+0.027*}

From Fig.~A1(f) MSX separates the two UCH\,{\sc ii} regions G10.46 and G10.47
45$''$ or 1.2\,pc apart, hitherto unresolved by IRAS (18056-1952). A third
mid-IR source lies to the south, alias IRAS 18056-1954. Hatchell et al. (2000)
present JCMT SCUBA images of this region, revealing a strong peak in the
850$\mu$m flux at the source A of Wood \& Churchwell (1989), with
G10.460+0.027A also weakly detected. Garay et al. (1993) show 20cm radio observations.

\subsubsection{[WC89] G10.623--0.384*}

MSX images of the environment of G10.623 reveal a complex of mid-IR sources, as
shown in Fig.~A2(a), of which the UCH\,{\sc ii} itself has a nearby, fainter, 
point source 25$''$ to the north east, plus a brighter mid-IR source
G10.598--0.383 to the south west 90$''$ (8 pc) away. The IRAS source
18075--1956 is dominated by the latter, at least in the 12--25$\mu$m
bands.

\subsubsection{[KCW94] G10.841--2.592}

Fig.~A2(b) shows that a bright IR source is coincident with the 
radio position of G10.841, with two close, fainter companions $\sim$30$''$ away
to the east and north-west. G10.841 itself, coincident with the GGD 27 complex, has
been the subject of  high resolution mid-IR imaging by Stecklum et al.
(1997). Peeters et al. (2002) have recently presented ISO spectroscopy of G10.841.


\subsubsection{[WC89] G11.938--0.616}

Fig.~A2(c) indicates that G11.938 is elongated, with a much fainter IR
source 70$''$ away (1.4 pc) to the SW.

\subsubsection{[WC89] G12.209--0.103*}

The UCH\,{\sc ii} G12.209 is a very weak mid-IR source according to
Fig.~A2(d), with two potentially related and much 
brighter IR sources that dominate the 21$\mu$m flux.  The closest of these is
60$''$ or 4\,pc away to the SW at G12.193--0.104.   Kim \& Koo (2001) present
low resolution 21cm VLA observations of G12.209, revealing  
a spatial morphology very similar to the mid-IR, with the two
sources to the SW and SE brighter than the UCH\,{\sc ii} itself, such that
these are not hot cores. Faint extended
radio emission also extends NE of the G12.209, as in the mid-IR.
Hatchell et al. (2000) 
present 850$\mu$m and 450$\mu$m SCUBA images of this region, showing
a peak associated with the UCH\,{\sc ii} region.

\subsubsection{[WC89] G12.429--0.049}

G12.429 appears single but is extremely weak at 12$\mu$m and 21$\mu$m as 
measured by MSX. 
It has the bluest color index of our sample, implying the highest dust
temperature. This object is well isolated with no other mid-IR source
within 10\,pc as illustrated in Fig.~A2(e). 
Kim \& Koo (2001) have presented 21cm radio data of G12.429.

\subsubsection{[KCW94] G18.146--0.284}

Fig.A2(f) reveals that 
G18.146 lies at the peak of a very extended ($\sim$2\,pc) region of dust
emission with multiple sources arranged along a NS axis.

\subsubsection{[WC89] G19.608--0.234}

G19.608 is spatially extended with a faint halo, according to
Fig.~A3(a), albeit no other nearby mid-IR companions.

\subsubsection{[WC89] G20.080--0.135*}

MSX images reveal G20.080 to be double, with a potentially related
companion only 36$''$ or 0.6 parsec to the south -- see Fig.~A3(b). The
companion is brighter than G20.080 at 12$\mu$m, though comparable
at 21$\mu$m, such that it may be stellar in origin. IRAS 18253-1130
contains both sources. Faint, nearby companions lie to the north and west.

\subsubsection{[WC89] G23.455--0.201*}

This UCH\,{\sc ii} is unique amongst our sample in that it is essentially
invisible  to MSX at {\it both} 12 and 21$\mu$m -- Fig.~A3(c).  
The strong IRAS source 18319--0834 commonly used to constrain the spectral
energy distribution of G23.455 is in fact located 80$''$ to the south, 
within a complex, extended region centred at G23.437--0.209. 
Consequently, we measure an
upper limit of $\sim$2Jy for G23.46 itself at 21$\mu$m, in contrast with 122\,Jy
obtained from the IRAS PSC at 25$\mu$m. We could find no evidence for an
error in the published radio position, or with our MSX coordinates. 
Indeed, Kim \& Koo (2001) present 21cm radio observations that confirm the
mid-IR view, namely that G23.455 lies at the northern edge of a strong 
radio source, spatially coincident with G23.437--0.209 which must be more
evolved than a hot core. Further to the south
radio emission extends along an east-west direction,  coincident
with a series of fainter mid-IR sources. 

\subsubsection{[WC89] G23.711+0.171*}

Fig.~A3(d) reveals that G23.711 is double, with a companion 36$''$ to
the south east. The companion has a comparable 12$\mu$m flux to the UCH\,{\sc ii},
but is 30\% fainter at 21$\mu$m which suggests that it may be stellar in origin. 
IRAS 18311--0809 contains both sources. Kim \& Koo (2001) included G23.711
in their 21cm survey of UCH\,{\sc ii} regions, also showing a spatial extension
from G23.711 to the SE.

\subsubsection{[WC89] G25.716+0.049*}

The UCH\,{\sc ii} region G25.716 appears to be single, albeit with a nearby brighter
mid-IR region 100$''$ or 4\,pc to the south -- see Fig.~A3(e).  
IRAS 18353--0628 is centred on the brighter IR source, 
such that an IR luminosity inferred from IRAS for the
UCH\,{\sc ii} would represent a strong overestimate. This UCH\,{\sc ii} region was
also studied at 21cm by Kim \& Koo (2001). Although the radio
spatial morphology closely matches the mid-IR view, Kim \& Koo argued that
the source to the south is not physically connected to G25.716 owing
to a very different radial velocity.

\subsubsection{[KCW94] G28.200---0.049}

G28.200 appears to be single from MSX 21$\mu$m images presented in Fig.~A3(f).

\subsubsection{[KCW94] G28.288--0.364}

MSX imaging reveals the core of G28.288 be elongated, as illustrated
in Fig.~A4(a).  There is also a companion a factor of four times 
fainter at 21$\mu$m 100$''$ to the east. IRAS 18416--0420 contains both sources.

\subsubsection{[WC89] G29.956--0.016}

G29.956 is the prototypical cometary UCH\,{\sc ii} region, and is amongst the brightest of our sample 
at mid-IR wavelengths despite its large distance (7.4 kpc, Churchwell et al. 1990). Fig.~A4(b) reveals 
the core of G29.956 to be extended. Very high spatial resolution (0.5$''$)
mid-IR imaging has recently been presented by De Buizer et al. (2002).
The closest mid-IR source is G29.935--0.055, an extended region
located 2.5$'$ away to the south east, which is also prominent in 
21cm radio observations of  Kim \& Koo (2001). These, together with
a further source to the SW, lie within a complex extended radio region.

This UCH\,{\sc ii} region is excited by an O5--6V star according to near-IR spectra 
discussed  by Watson \& Hanson (1997) and Hanson et al. (2002). Peeters
et al. (2002) present the ISO spectrum of G29.956, which is analysed
by Morisset et al. (2002) with reference to the ionizing source (see, however, Lumsden et al. 2003).

\subsubsection{[WC89] G30.535+0.021}

G30.535 
appears to have a single core in the mid-IR as shown in Fig.~A4(c), albeit with a faint
halo.

\subsubsection{[WC89] G31.414+0.310}

G31.414 appears to have a single core, again with an  extended halo -- see Fig.~A4(d).
 Garay et al. (1993) present 2-20cm radio images of G31.414.
The SCUBA sub-mm emission reveals a 
central peak coincident with the UCH\,{\sc ii} region,
with some extension to the south (Hatchell et al. 2000).

\subsubsection{[KCW94] G32.798+0.190}

G32.798 is single according to our MSX images presented in Fig.~A4(e). Peeters
et al. (2002) presented the ISO spectrum of G32.80. Garay et al.
(1993) present 2-20cm VLA radio observations, whilst Kurtz et al. (1999)
have identified extended emission from 3.6cm imaging.

\subsubsection{[WC89] G33.915+0.110}

G33.915 again appears to have a single core, but with a faint halo --
Fig.~A4(f). The ISO spectrum of G33.92+0.11 is
presented by Peeters et al. (2002). Fey et al. (1992) and Garay
et al. (1993) discuss the radio morphology of G33.915 from 2 to 20cm.

\subsubsection{[WC89] G34.255+0.145}

G34.255 appears fuzzy with a strong dust tail extending east-west
below the central source -- Fig~A5(a). G32.455 is a well studied
cometary UH\two\ region at radio wavelengths (e.g. Fey et al. 1992).

\subsubsection{[WC89] G35.199--1.743}

G35.199 contains a central dense, core but with an envelope with a condensation
extending 90$''$  or $\approx$ 2 pc to the south -- see Fig.~A5(b). Takahashi et 
al.
(2000) infer stellar properties of G35.199 from mid-IR imaging in the [Ne\,{\sc ii}] 
12.8$\mu$m filter.

\subsubsection{[WC89] G37.545--0.112}

G37.545 appears spatially extended with another possible embedded source --
Fig.~A5(c). Kim \& Koo (2001) present 21cm radio observations of G37.545.

\subsubsection{[KCW94] G37.874--0.399}

G37.87 appears to be single, with faint extended dust emission extending
$\sim$60$''$, as illustrated in Fig.~A5(d).

\subsubsection{[WC89] G43.889--0.783}

This UCH\,{\sc ii} appears as a point source at 21$\mu$m according to Fig.~A5(e),
with extended emission
extending SE containing a condensation. The sub-mm SCUBA image of Hatchell et al.
(2000) reveals a single peak towards to the radio position.

\subsubsection{[WC89] G45.071+0.132 and G45.122+0.132}

G45.071 and G45.122 are 3$'$ apart on the sky, such that their physical
separation is $\sim$6 pc if they lie at a common distance of 6.5\,kpc.
Churchwell et al. (1990) suggest a distance of 6\,kpc to G45.07 and
6.9\,kpc to G45.12. These UCH\,{\sc ii} regions have previously been resolved by
IRAS at 12 and 25$\mu$m, such that G45.071 is IRAS 19110+1045, whilst
G45.122 is IRAS 19111+1048. The latter has a companion 50$''$ away to the
north west, at G45.134+0.144 -- see Fig.~A5(f). This weakly contaminates the 25$\mu$m IRAS 
measurement of G45.122. Lumsden et al. (2003) infer a cluster of OB
stars, rather than a single O star, power both UCH\,{\sc ii} regions, while
Takahashi et al. (2000) infer stellar properties  of G45.122 
from mid-IR imaging in [Ne\,{\sc ii}] 12.8$\mu$m. Hunter  et al. (1997) discuss 
sub-mm observations of the molecular cores containing these two UCH\,{\sc ii} regions,
suggesting that G45.122+0.132 is at a more advanced state of star formation.

\subsubsection{[WC89] G45.456+0.060 and G45.466+0.046*}

G45.456 is very bright at mid-IR wavelengths, with an elongated core.
Of the three nearby, much fainter 21$\mu$m sources seen in Fig.~A6(a), 
only the object 60$''$ to the
north west is also seen at 12$\mu$m. Indeed, the faint source 70$''$ to
the east of G45.456 is the UCH\,{\sc ii} G45.466+0.046 which has an exceptionally
red mid-IR color of F(21$\mu$m/12$\mu$m)$>$30. If these lie at the same
distance their separation is 2\,pc. A further bright sources lies 4.5$'$ away
to the NW at G45.479+0.133. Another H\two\ region, which is very bright at 
21$\mu$m,
lies to the NW at G45.479+0.133, and is also thought to be part of the same star
forming complex. 

Feldt et al. (1998) discuss high spatial resolution near- and mid-IR
observations of G45.456, whilst Garay et al. (1993) present 6 and 20cm radio maps.
Lumsden et al. (2003) have recently argued in favour of a 
cluster of OB stars ionizing the UCH\,{\sc ii} region, instead of a single star. 


\subsubsection{[KCW94] G48.606+0.024*} 

This UCH\,{\sc ii} region is revealed as a point source within a large
extended mid-IR emitting region, with a nearby bright source centred
on G48.595+0.044 -- see Fig.~A6(b). The IRAS source 19181+1349 
is dominated by the latter object. Another faint source lies
nearby to the north. 3.6cm imaging by Kurtz et al. (1999) revealed
extended emission in the H\two\ region surrounding G48.606.

\subsubsection{[WC89] G49.490--0.370* (W51d)}

This is an extremely complicated portion of the well known GH\two\ region 
W51
at mid-IR wavelengths, as
indicated by Fig.~A6(c). 
G49.490 is an exceptionally bright source at
21$\mu$m. Another mid-IR bright source lies very close, 40$''$ ($\approx$ 2 pc) away,
to the south east at G49.488--0.381, that possesses a bright dust tail. IRAS
19213+1424 includes both these bright sources, such that the IR luminosity
of the UCH\,{\sc ii} region will be strongly overestimated via use of the IRAS flux.
Consequently for this particular example, arguments made by Lumsden et al. (2003) 
in favour of a  {\it cluster} of OB stars powering this UCH\,{\sc ii} region will
be affected. Several other fainter mid-IR sources lie to the south, south west, west and north east.
High resolution mid-IR imaging and spectroscopy of W51d, alias W51 IRS2,
have  been presented by Okamoto et al. (2001) and by Kraemer et al. (2001). 

\subsubsection{[WC89] G54.094--0.060}

This is an isolated double source, with a dust tail extending
to the south east -- see Fig.~A6(d).

\subsubsection{[KCW94] G60.884-0.128}

This source, alias Sh~2-87,  sits within an small extended region with a nearby (1 pc if
physically related) point source to the west at G60.872--0.107 -- see Fig.~A6(e).
Peeters et al. (2002) present the ISO spectrum of G60.884. Kurtz et al. (1999)
identified extended radio emission at 3.6cm in G60.884.

\subsubsection{[WC89] G61.473+0.093}

G61.473 (Sh~2-88) appears to have an spatially extended core, plus an extensive dust
halo with a diameter of 80$''$ ($\sim$2~pc) -- see Fig.~A6(f). 
6cm and 20cm radio observations have been presented by Garay et al. (1993).

\subsubsection{[KCW89] G70.293+1.600* and G70.330+1.586*}

Fig.~A7(a) reveals that G70.293, alias K3-50A, lies at the centre of 
a small cluster  of at least three sources, of which it is the brightest 
mid-IR source.  G70.330 is much fainter and lies $2.5'$ away to the NE.
All sources contribute to the IRAS source 19598+3324.

Okamoto et al. (2003) have recently presented mid-IR 
spectroscopy and imaging of G70.293 suggesting that it is ionized by
a cluster of two or three late O stars, instead of a single dominant 
source. Lumsden et al. (2003) arrives at similar conclusions from near-IR
observations.

\subsubsection{[WC89] G75.783+0.343*}

The relatively high spatial resolution obtained by MSX permits the UCH\,{\sc ii}
G75.783 to be identified as a rather weak mid-IR source, in contrast to the
commonly assumed coincidence between G75.78 and IRAS 20198+3716 -- see
Fig.~A7(b). The IRAS flux, at least for the 12--25$\mu$m bands, is
totally dominated by a source 60$''$, or 1.2\,pc to the south west. 
The IRAS source has a 12$\mu$m flux of 423\,Jy of which only 8\,Jy 
is from G78.78 3 according to our Band~C MSX images. 

\subsubsection{[WC89] G75.835+0.400}

G75.835 (Sh~2-105), located 5 arcmin north of G75.783 is a bright mid-IR
source, dominating IRAS 19111+1048. It has a close, much fainter 
companion 45$''$ to the north -- see Fig.~A7(c). If these lie at the same distance, as
would appear probable, this corresponds to a separation of 1.2\,pc.
Garay et al. (1993) present 2-20cm radio maps of G75.835.

\subsubsection{[KCW89] G76.383--0.621}

This UCH\,{\sc ii}, alias Sh~2-106, appears to be single, though is spatially extended some 90$''$
in the north-south direction. A faint IR point source is seen 80$''$
($\approx$ 0.4 pc) to the south in Fig.~A7(d). Kurtz et al. (1999) identified extended
radio emission in G76.383.

\subsubsection{[KCW94] G78.438+2.659}

Fig.~A7(e) indicates that G78.438 appears single, with a very faint dust tail to the south, and no
nearby mid-IR source. Once again, Kurtz et al. (1999) observed extended radio emission at 3.6cm
for G78.438.

\subsubsection{[KCW94] G81.679+0.537 and G81.683+0.541 (DR21)}

Mid-IR imaging of DR21 reveals a 21$\mu$m bright extended source,
containing G81.679+0.537 and G81.683+0.541, though centred on neither.
In addition, a diffuse region
extends east--west -- see Fig.~A7(f). Several faint point sources are also
seen, the brightest of which is 3$'$ to the NW. Kraemer et al. (2001)
present high resolution mid-IR imaging of DR21, revealing separate N and SW
components at 13$\mu$m, together with comparisons to radio continuum measurements. No IRAS
source is catalogued at the position of DR21. Peeters et al. (2002) present
a mid-IR spectrum obtained with ISO.

\subsubsection{[KCW94] G109.871+2.113}

G109.871 is a bright, though extended, 21$\mu$m source, as shown in
Fig.~A8(a), with a
faint companion 2$'$ ($\approx$ 1 pc) north west. G109.871 is extremely faint
at 12$\mu$m such that it has a 21/12$\mu$m ratio of $\sim$47, the highest
of our present sample of UCH\,{\sc ii} regions, indicating the lowest dust temperature.
Unusually, no extended radio emission was observed by Kurtz et al. (1999).

\subsubsection{[KCW94] G111.282--0.663}

Fig.~A8(b) reveals G111.282 (Sh~2-157B) to lie at the centre of an extended 2$'$
complex, with faint emission to the NE. In addition, a companion lies
several arcmin to the south at G111.279--0.707 that is extremely red, since it is barely 
detectable at 12$\mu$m. No  stellar photospheric
features were detected by Hanson et al. (2002) from NIR spectroscopy
of G111.282. Kurtz et al. (1999) identified extended ratio emission surrounding this
UCH\,{\sc ii} region.

\subsubsection{[KCW94] G111.612+0.374}

G111.61, alias Sh~2-157, is identified as a bright, spatially extended source with no nearby mid-IR
companions in Fig.A8(c). Hanson et al. (2002) present NIR spectroscopy of the
central source, revealing strong nebular HeI 2.11$\mu$m emission,
indicating a very hot (early O) ionizing source.
Peeters et al. (2002) discuss the ISO spectrum of G111.612. No extended radio
emission was observed in this UCH\,{\sc ii} region by Kurtz et al. (1999).

\subsubsection{[KCW94] G133.947+1.064}

This UCH\,{\sc ii} region, alias W3(OH), appears single at mid-IR wavelengths --
see Fig.~A8(d) -- with a halo and a nearby faint 
companion $\sim$30$''$ to the east. High resolution mid-IR imaging was presented
recently by Stecklum et al. (2002).

\subsubsection{[KCW94] G139.909+0.197}

G139.909 appears to be a point source at 21$\mu$m, Fig.~A8(e), but has
a close faint 12$\mu$m companion, which is probably a star. Hanson et
al. (2002) present NIR spectroscopy of (probably) the central star of G139.909,
revealing a late O or early B spectral type.

\subsubsection{[KCW94] G192.584-0.041*}

This UCH\,{\sc ii} region is extremely faint at 21$\mu$m relative to a nearby
extended source, $1'$ to the south which dominates the
IRAF 12 and 25$\mu$m fluxes -- see Fig.~A8(f).
 Indeed, G192.584 is not detected with MSX at 12$\mu$m, such
that it has a 21/12$\mu$m flux ratio in excess of $>$11. 
Other faint, equally red, sources lie to the south and to the 
west which are probably other members of the same complex.

\subsection{Overall morphology}

We have listed in Table~1 the UCH\,{\sc ii} radio morphology\footnote{Radio 
morphologies are listed as either cometary,
core-halo, spherical or shell by Wood \& Churchwell (1989)} 
as found by the radio measurements at spatial scales of
a few arcsec. We also indicate in this table our evaluation of the dust
morphology as indicated by the MSX 21$\mu$m images, which is on spatial
scales of a few arcmin or more. We call sources {\em single} if the core
image appears roughly circular with little or no elongation. Those labelled
{\em extended} are elongated or have fainter dust surrounding the core
which may contain other objects. Sources with multiple cores, or with
``companions'' nearby are called {\em double}, or {\em multiple}.  There
does not seem to be any correlation between the radio and the dust
morphologies, which may not be too surprising given the very different
scales of these phenomena.

Many of the sources appear to be resolved by MSX, i.e. their diameters
are $\geq 18''$, the spatial resolution of the instrument. This resolution
corresponds to different physical scales, depending on the source
distances, as may be seen by examination of Figures~A1--8.
Looking at the two nearest  sources, their core diameters correspond to a few
tenths of a pc. Many others have larger core diameters of up to $\approx$
1 pc.  The surrounding halos extend to even greater distances in some
cases.

We were surprised to find most of our MSX images to have MIR sources in 
addition to the UCH\,{\sc ii} region located nearby. Those we consider double or
multiple have companions within about 5 pc (projected) if they are at the
same distance (radial velocities are generally not available). In some
cases, already noted in the discussion of individual objects, the IRAS photometry
will be dominated by the companion.  At 21$\mu$m normal stars will be
fainter than at 12$\mu$m, so we expect the companions to be dust emission
sources. Some might be `hot cores' (e.g., Churchwell 2002), although 
Kim \& Koo (2001) revealed that many such companions are radio sources, 
indicating that they are at a more advanced state of evolution. 
There are 53 sources in our Table 1, of which 12 are single with no
extended dust or obvious companions. It appears that in most cases, this
sample of {\em field UCH\,{\sc ii} objects are not isolated star formation sites
but are connected with nearby (few pc distant) activity}.

\section{Comparison between MIR--FIR and radio fluxes}\label{comp}

\subsection{The standard model}

A well known model of the dust emission from an UCH\,{\sc ii} region (e.g., Churchwell 1999) is
spherically symmetric (1D) with two major constituents: the inner ionized
hydrogen volume and an outer thick shell of molecular gas and dust. The
outer boundary of the ionized hydrogen region is typically up to 0.1 pc in
radius, the surrounding cocoon dust some ten times larger.  The cocoon is 
optically thick to the visible--UV radiation of the exciting star and is 
thus heated by it (Kahn 1974; Osorio et al. 1999). 

The dust emission will dominate the observed spectral energy distribution
(SED). This typically has a broad peak, close to 100$\mu$m, fortuitously coincident
with the long wavelength IRAS filter. Thermal radiation from the dust cocoon
extends from about 1 mm  down to a few microns. This dust is
{\em not} at a single temperature but has a broader wavelength
distribution than a black body. Most of the dust is at a temperature of
about 30K, although some of that nearest the star could be at 100K or
more (see discussion in Wolfire \& Churchwell 1994).  Their modelling of
the IRAS SEDs indicated that the dust density distribution in the shell
was more or less constant (no fall off with radius). A more recent study
by Hatchell et al. (2000) using the {\sc dusty} radiative transfer code required
a $r^{-3/2}$ density profile distribution to reproduce the sub-mm radial
emission profiles. 2D geometries 
in the near future will provide a better model for comparison purposes.  
We use the phrase `standard model' to refer to the
situation where the radio emission comes from an inner spherical region
and the IR emission from the outer cocoon.

Radiation from wavelengths longer than about 1 mm will arise from the
inner region of ionized hydrogen surrounding the central OB star.  Lyman
continuum photons are emitted from the stellar surface and ionize the
hydrogen. According to the standard model, the H\two\ region is {\em
ionization} bounded, that is, all the LyC photons are used up within this
volume which produces the radio emission. Note that some photons
emitted by the star might be absorbed by dust within the H\two\ 
region\footnote{
Lyman $\alpha$ photons are efficiently absorbed by dust in compact
H\two\ regions and probably represents the dominant dust heating 
mechanism}. 
These photons would not produce radio emission and would not be counted
when one is trying to infer properties of the star from the H\two\ region.
Whether this is an important problem for UCH\,{\sc ii} regions will be discussed
below. 
 
We are interested here in the relationship between the IR emission from
the dust, which is determined by the stellar luminosity which for hot
stars is the UV radiation, and the radio emission of the H\two\ region, 
which
is dependent upon the EUV luminosity of the star.  This ratio ought to be
spectral type dependent, ranging from $10^3$ for early O type stars to
$10^6$ for early B types (Churchwell 1999).  We will now see if the standard
model of UCH\,{\sc ii} regions can fit stars with this wide a range of parameters.

\subsection{Comparison of MSX 21$\mu$m flux with 25$\mu$m IRAS flux}

\begin{figure} 
\epsfxsize=8.8cm\epsfbox[0 400 504 790]{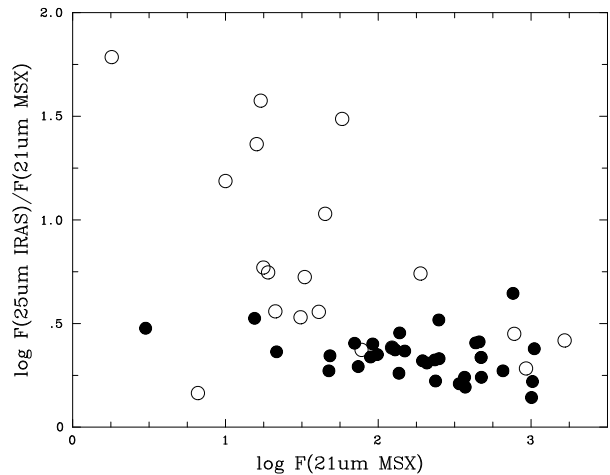}     
\caption{Comparison between $30''$ aperture 25$\mu$m IRAS fluxes
for Galactic UCH\,{\sc ii} regions and the $18''$  aperture 21$\mu$m MSX fluxes,
based on data from  Wood \& Churchwell (1989) and 
Kurtz et al. (1994). Open circles: sources with brighter close companions; 
filled circles: sources without such companions. 
Agreement between the 21$\mu$m and 25$\mu$m  measurements
for sources without companions is good.}
\label{fig9} 
\end{figure} 

We have identified a number of UCH\,{\sc ii} regions which at the spatial
resolution of MSX are found to have nearby companions. These may be
affecting the IRAS Point Source Catalog (PSC) entries, particularly the
long wavelength $100\mu$m filter with its $2'$ spatial resolution. In some
cases the companion may dominate the flux, even at the shorter
wavelengths. As we wish to infer the parameters of the dust cocoons from
the MIR--FIR photometry, let us first examine quantitatively how much of a
problem contamination plays.

In Figure~\ref{fig9} we plot the 25$\mu$m/21$\mu$m (log) flux ratio vs. the
21$\mu$m (log) flux (both are distance independent), where we have
distinguished between sources without bright companions and those with
them. Many of the latter have wildly different fluxes in the MSX and IRAS
filters. Leaving these outliers aside, there is no trend of flux ratio
with the strength of the MIR flux, and the mean ratio (not log units) is
2.20$\pm$0.46. The differences in the IRAS and MSX filter bandwidth alone
accounts for a factor of 1.6 (recall Fig.~\ref{fig0}). The rest is due to
the IRAS filter being somewhat redwards of the MSX filter, thus nominally brighter
in these red sources. In the following figures we shall concentrate on 
those objects without bright companions (filled circles) although we will 
show all sources throughout.
  
One further note of caution is that for our comparison between MSX and radio
fluxes, we have limited the mid-IR flux to the spatial resolution of
the instrument, such that the total mid-IR flux of extended sources is
underestimated. One outlier in Fig.~\ref{fig9} without a close companion, 
G5.972--1.174, is highly extended,
such that our quoted MSX 21$\mu$m flux of 762 Jy, increases dramatically
for larger apertures, i.e. 1600 Jy (36 arcsec radius) or
2400 Jy (60 arcsec radius). Consequently, the quoted 25$\mu$m/21$\mu$m
flux of 2.4 is highly aperture sensitive for this one case especially, 
decreasing to 1.15 with an adopted 36 arcsec radius, or 0.77 for a one arcminute MSX radius. 

\subsection{Color-magnitude diagrams}

\begin{figure} 
\epsfxsize=8.8cm\epsfbox[0 400 504 790]{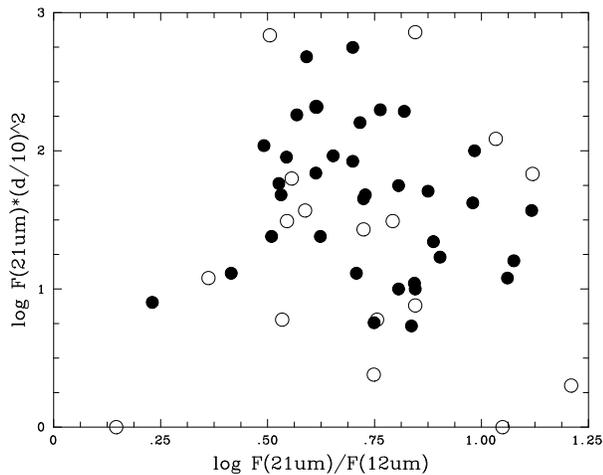}
\caption{IR colour-magnitude diagram of UCH\,{\sc ii} regions:  Comparison between
MIR colour (21$\mu$m/12$\mu$m) and 21$\mu$m flux scaled to a uniform
distance of 10 kpc. As above, we distinguish between those sources with 
bright
companions and those without. In this figure there is essentially no
distinction.}
\label{fig10}
\end{figure} 

Stellar astrophysicist's typically prepare colour magnitude diagrams
(CMD), which plot the luminosities of stars vs. their colors, or
temperature. As we are not observing the exciting stars of the UCH\,{\sc ii}
regions directly, we have to use other measurable parameters. The MIR--FIR
fluxes, corrected to a uniform 10 kpc distance, may serve as a surrogate
for the luminosity since nearly all of the stellar radiation has been
converted to heating the dust cocoon.  We can measure the color of the
cocoon, an indicator of the dust temperature distribution, by obtaining
the MSX 21$\mu$m/12$\mu$m flux ratio.

Fig.~\ref{fig10} presents an MIR CMD, namely the MSX 21/12$\mu$m index versus
the 21$\mu$m flux, adjusted to a uniform distance of 10 kpc.  We see a
fairly wide scatter along the abscissa suggesting a large range of colours
for the dust cocoons. This indicates that they do not all have the same
slopes of the SED on the short wavelength side. The vertical extent, of
course, represents different luminosities of the dust, thus of the central
exciting stars. We see that there is little if any dependence of the
luminosity (ordinate) on the color.  Thus the dust luminosity and its SED
are not correlated.  The color of the dust cocoon will have to be
accounted for in detailed modeling of individual sources. 

We have considered
whether effects other than temperature might affect the 
21/12$\mu$m index, namely
nebular contributions and extinction. 
Many sources are strong emitters in mid-IR
fine structure lines of neon and sulphur, of 
which [Ne\,{\sc ii}] 12.8$\mu$m and
[S\,{\sc iii}] 18.7$\mu$m lie in the 12 and 21$\mu$m filters. 
From inspection of
a range of UCH\,{\sc ii} regions observed with ISO/SWS (e.g. G29.956--0.016
in Fig.~\ref{fig0}), we conclude that this contribution at most 10\%
of the continuum dust emission. Variable extinction may 
be more probematic, since many sources are very heavily reddened. The 
12$\mu$m Band C filter
of MSX does include the red wing of the familiar silicate absorption feature 
centred at 9.7$\mu$m (e.g. Fig~2 of Morris et al. 2000), this is compensated to some
extent by the much greater linewidth of the 21$\mu$m Band E filter. Nevertheless, 
one should bear in mind that 21$\mu$m and especially 12$\mu$m flux measurements are
affected to some degree by extinction, in contrast with far-IR  and radio fluxes.

\subsection{MIR vs. radio relationships}

Radio fluxes from Wood \& Churchwell (1989) and Kurtz et al. (1994) were
generally obtained at high spatial resolution with interferometric
techniques, and so are restricted to the central regions of
each UCH\,{\sc ii} region at 0.5--5$''$ resolution.  In contrast, the somewhat
lower spatial resolution of the MSX fluxes presented here sample a
larger region.  There is increasing evidence in the literature
that some LyC photons are finding their way out beyond the UCH\,{\sc ii} region where
they ionize the hydrogen found there. 

These halos have been investigated by Kurtz et al. (1999) and Kim \& Koo (2001) 
who re-examined the radio properties of UCH\,{\sc ii} regions at larger
scales (see also Araya et al. 2002). As we have discussed above, there
is an excellent consistency between the Kim \& Koo 21cm dataset
and our mid-IR MSX images, particularly for complex regions such as
G5.885--0.392, G10.304--0.147, G12.209--0.103 and G23.455--0.201.
In most cases, the
larger aperture leads to a higher radio flux, i.e. the commonly adopted
radio fluxes (dense, ionized components) provide only a fraction of the
total when the less dense, extended surrounding regions are considered.
These extended regions range in size from 15$''$ to more typically 60$''$.
Larger aperture fluxes generally exceed the Wood \& Churchwell (1989)
and Kurtz et al. (1994) VLA results at 6cm by
a factor of 3, although the factor may be as large as 29 (for G60.88--0.13;
Araya et al. 2002).
 
Unfortunately, the sample observed with these
techniques is small to date, so we have to follow the original VLA fluxes from
Table~1 for our analysis of the UCH\,{\sc ii} region sample. Nevertheless,
we will keep this multiplicative factor in mind in what follows,
hoping that it is reasonably the same from source to source. The existence
of ionized hydrogen halos of UCH\,{\sc ii} regions implies that the standard model
is {\em density}, not {\em ionization}, bounded.

\begin{figure} 
\epsfxsize=8.8cm\epsfbox[0 400 504 790]{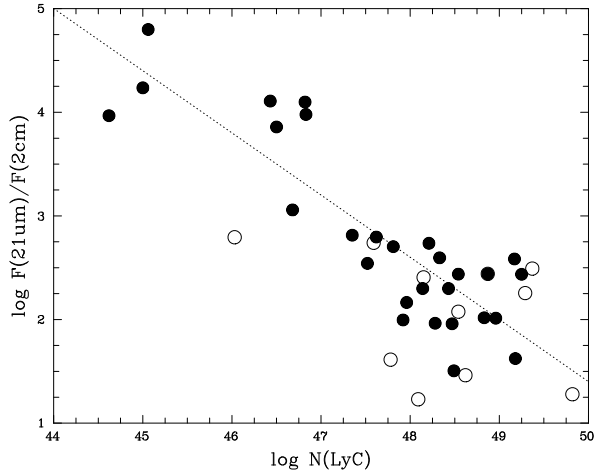}
\caption{Comparison between the MIR/radio (21$\mu$m/VLA 2cm) index of
Galactic UCH\,{\sc ii} regions versus ionizing fluxes, again
discriminating between sources with bright companions (open 
circles) and those without (filled circles). A linear fit
to the latter is shown as a dotted line (see text).}\label{fig11}
\end{figure} 

We ratio the 21$\mu$m and VLA B-array 2cm fluxes as a function of the
observed N(LyC) photons in Fig.~\ref{fig11} for all UCH\,{\sc ii} regions for
which 2cm fluxes are available.  The abscissa, which is distant dependent,
is used as a surrogate for the temperature (spectral type) of the exciting
star.  We see a trend in the MIR/radio flux ratio with the temperature of
the exciting star, which is predicted by the standard model. A linear
regression analysis leads to
\[\log(21\mu{\rm m}/2{\rm cm}) = C_1 (50-\log{\rm N(LyC)})  + C_2,\] 
where $C_1 = 0.600\pm 0.068$ and $C_2 = 1.401 \pm 0.178$.
This is an empirical relationship which will be useful below for it's 
scatter. A slight note of caution is necessary, however, since
the 21$\mu$m/2\,cm ratio would
be suppressed in the case of extremely high extinction, because the denominator
has no dependence, whilst the numerator has a weak dependence -- recall
that a visual extinction of 100 magnitudes corresponds to a 21$\mu$m extinction of 
0.5~mag (0.2 dex in Fig.~\ref{fig11}).

Kurtz et al. (1994) made an analogous plot of IR to radio flux ratio but
used the 100$\mu$m luminosity rather than the NLyC photons. They
attributed their scatter to two sources: 1) some LyC photons are being
absorbed by the dust and 2) some UCH\,{\sc ii} regions have multiple sources
within them.  Either or both of these effects could contribute to our
scatter in the following ways: 1) If LyC photons are being absorbed by the
dust, a point in Fig.~\ref{fig11} would have moved leftwards and upwards
in the diagram; 2) If other stars are present and contributing to the IR
flux, a point would have moved vertically upwards.  Before attributing 
our scatter to these same influences we realized that here might be an
additional parameter influencing the scatter in our plot.

The 21$\mu$m wavelength has a strong color term (recall Fig.~\ref{fig10}) due to
it being on the short wavelength slope of the SED, which is itself not
uniform from object to object.  We will next show an analogous plot using
the IRAS 100$\mu$m flux, which is at the peak of the dust cocoon SED,
taking account of the problem of the multiplicity of the sources.

\subsection{FIR vs. radio relationships}

\begin{figure} 
\epsfxsize=8.8cm\epsfbox[0 400 504 790]{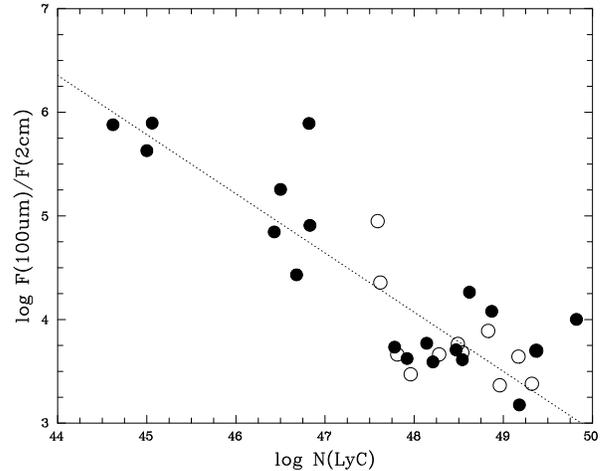} \caption{Comparison
between FIR /radio (IRAS 100$\mu$m /VLA 2cm) index of Galactic UCH\,{\sc ii}
regions versus ionizing fluxes, for sources with bright companions (open 
circles) and those without (filled circles). A linear fit to the latter
is shown as a dotted line (see text).}
\label{fig12}
\end{figure} 

In Fig.~\ref{fig12} we plot the IRAS 100$\mu$m flux/VLA B-array 2 cm flux vs.  
the N(LyC) photons.  We see a very clear linear relationship,
extending over 5 orders of magnitude in the number of N(LyC) photons 
and 3 orders in the IR/radio ratio.  
A fit to the UCH\,{\sc ii} regions without bright companions is shown as a
dotted line in Fig.~\ref{fig12}. The empirical regression relation is 
\[
\log (100\mu {\rm m}/2 {\rm cm}) = C_3 (50 - \log {\rm N(LyC)} ) + C_4,\]
where $C_3 = 0.571 \pm 0.048$ and $C_4 = 2.929 \pm 0.124$, having omitted 
the one outlier, G18.15--0.28, which lies well above the mean relation. 
(This source sits in a very extended region of dust emission with multiple fainter sources which may 
be affecting its fluxes - see  Fig.~A2). Notice that the 
scatter here is significantly less than that shown in Fig.~\ref{fig11}, when we 
used the 21$\mu$m wavelength. The scatter in this diagram presumably 
represents the effects of dust absorption of LyC photons and that of other 
stellar contributers to the dust emission.  These effects are relatively 
small. 

\begin{figure} 
\epsfxsize=8.8cm\epsfbox[0 400 504 790]{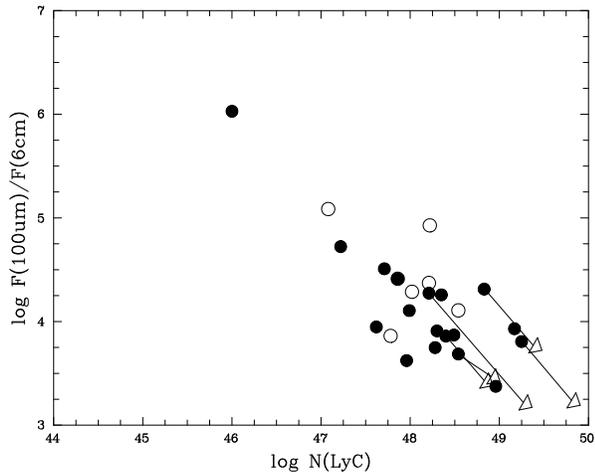} 
\caption{Comparison between FIR /radio (IRAS 100$\mu$m /VLA 6cm) 
index of Galactic UCH\,{\sc ii} regions versus ionizing fluxes, from Wood \& Churchwell 
(1989) with symbols as before. Arrows indicate revised 6cm and ionizing fluxes
using larger aperture Arecibo observations (Araya et al. 2002).}
\label{fig13}
\end{figure} 

Since many Wood \& Churchwell (1989) UCH\,{\sc ii} sources were without 2cm flux
measurements, in Fig.~\ref{fig13} we compare the IRAS 100$\mu$m to VLA 6cm index with Lyman
continuum ionizing flux. This relationship is similar
to that shown in the previous figure, except that there are no fainter
stars in the Wood \& Churchwell sample. 
Arrows indicate revised locations for those UCH\,{\sc ii}
observed (or inferred) with larger radio beams (Araya et al. 2002).
Notice that these more or less move {\em along} the empirical relationship 
towards higher radio fluxes and N(LyC), in keeping with predictions.

\begin{figure} 
\epsfxsize=8.8cm\epsfbox[0 400 504 790]{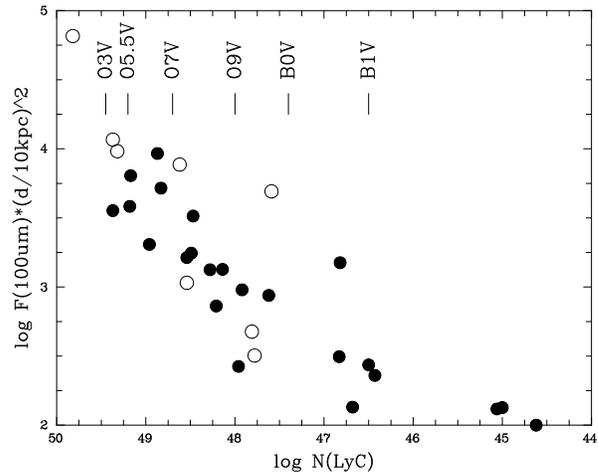}
\caption{Comparison between Lyman continuum flux inferred from radio and
IRAS 100$\mu$m flux scaled to a uniform distance of 10 kpc. Again, we
distinguish between those sources with (open) and without (filled)
bright companions.  We can interpret this diagram as the ZAMS for hot, luminous stars,
and so indicate typical ionizing fluxes for OB dwarfs (adapted from Smith et al. 2002,
following the revised dwarf temperature scale of Martins et al. 2002).}
\label{fig13a}
\end{figure}

Finally, let us construct a CMD for the exciting central stars of UCH\,{\sc ii}
regions, using as the ordinate the FIR luminosity, and as abscissa, the
N(LyC) photons as a surrogate for the spectral type. This result is shown 
in
Figure~\ref{fig13a}, and we reversed the abscissa so it looks like those 
of stellar astrophysicists.  Considering only the stars unaffected by 
companions, we see that there is a range in luminosity of 5 magnitudes 
between early O stars N(LyC) photons with $\approx~10^{49.5}$ s$^{-1}$) and 
early B  stars, with 
N(LyC) photons $\approx~10^{45}$ s$^{-1}$). This is broadly in agreement with 
models of hot, luminous stars as indicated in the key in the Figure. This
follows the recent OB grid of Smith et al. (2002), adapted to take effect of the revised
temperature calibration of Martins, Schaerer \& Hillier (2002).
We have not converted the FIR luminosity to a magnitude 
system on the figure, since the 100$\mu$m flux does not take into account dust 
emission from longer wavelengths.  This correction is a factor of a few. 
Similarly, the N(LyC) photons measured for the inner UCH\,{\sc ii} region have not 
accounted for the halo of ionized material.  We already saw that this 
correction was also a factor of a few.

We see in Figure~\ref{fig13a} that the objects without much influence of a
companion have a reasonably tight relationship between their luminosities
and temperatures.  We can interpret this diagram as the ZAMS for hot,
luminous stars.  There is not much scatter although G18.15--0.28
(recall Figure~\ref{fig12}) again stands out above the relation.  It could be that
this UCH\,{\sc ii} region has several close companions which contribute to the luminosity
of the dust, but not to the N(LyC) photons.

\section{Conclusions}\label{dis}

Young, massive stars are obscured at UV, optical and near-IR wavelengths due to extremely
high extinction from their circumstellar, birth envelopes. For every magnitude of 
extinction suffered in the mid-IR at 25$\mu$m, that in the K-band is 20 times greater,
whilst the visual extinction is 200 times greater (e.g. Fig~2 of Morris 
et al. 2000). Consequently, only 10\% of OB
stars ionizing an UCH\,{\sc ii} region can be detected in the near-IR (Hanson et al. 2002).
Since the star may not be directly observed in most cases, we have to rely on indirect
probes at mid-IR, far-IR and radio wavelengths. In the current study we 
have presented MSX mid-IR observations of dust surrounding UCH\,{\sc ii} regions.
These are sites of massive star birth, with central ionizing stars spanning
 spectral types of early B in G109.871+2.113 to early O in G10.623--0.384, 
at a higher spatial resolution than that which IRAS offered. Only
25\% of all radio selected UCH\,{\sc ii} regions surveyed here were {\it not} 
found to be extended, or have close companions, themselves often radio emitters.
In most cases, field UCH\,{\sc ii} regions lie within larger complexes of star formation.
Consequently, previous use of IRAS fluxes for the spectral energy distributions of
UCH\,{\sc ii} regions would have led to strong overestimates of 
bolometric luminosities in $\sim$50\% of cases. This result, although 
widely anticipated, reduces inferred stellar luminosities,  {\it without} 
affecting LyC fluxes. This may affect recent claims that clusters of
stars, rather than a single dominant source, ionize individual UCH\,{\sc ii} 
regions (e.g. G49.490--0.370, Lumsden et al. 2003). 

Comparison between MSX 21$\mu$m and 12$\mu$m fluxes permit
colour information on the dust cocoon to be made in each case. We find a large range
of colours, indicating a variety of dust temperature distributions. 
Consequently, because of this colour term,
we may not blindly use 21$\mu$m fluxes as representative of the dominant 
far-IR luminosity, and so resort to the IRAS 100$\mu$m fluxes for
those UCH\,{\sc ii} regions  that appear to well isolated.  
From a comparison between MSX 21$\mu$m and  IRAS 25$\mu$m fluxes, 
we identify those
IRAS sources for which the UCH\,{\sc ii} regions are the sole or principal source of mid-IR radiation,
and so the likely principal source of far-IR flux. A comparison is made between the
100$\mu$m flux and Lyman continuum radiation, inferred from radio observations, revealing
a linear relationship, as predicted by the standard model for UCH\,{\sc ii} regions. 
Any remaining
scatter may be attributed to dust absorption by the emitted LyC radiation and fainter
companions within the UCH\,{\sc ii} region. Finally, we compare the UCH\,{\sc ii} region 
Lyman continuum flux with observed 100$\mu$m fluxes, adjusted to a uniform distance,
again revealing a tight spectral type dependence, also in general
accord with the standard model.

Overall our results are encouraging, but we should not neglect remaining problems. As
discussed above, MSX reveals large scale information on the stellar nurseries of massive stars, 
with a spatial resolution somewhat higher than IRAS. In contrast, ground-based near-IR and
mid-IR observations at a much better spatial resolution reveal structure on a quite
different, complementary scale.
Most radio observations compare more closely with the latter, except for
recent data of Kurtz et al. (1999) and Kim \& Koo (2001). In the future, 2MASS
and SIRTF  will permit comparisons of large numbers of UCH\,{\sc ii} regions to be made on a similar scale.
 Individual massive Young Stellar Objects (YSOs) have
recently been studied using MSX by Lumsden et al. (2002). In contrast, our next study will 
investigate the mid-IR properties of a complete radio sample of 
Giant H\two\ (GH\two) regions in the Milky Way.


\section*{Acknowledgements}  

PAC and PSC appreciate continuing support by the Royal Society and the
NSF, respectively. We wish to thank Martin Cohen for help with MSX flux
calibration, Robert Stencel, Nathan Smith and Matt Redman for useful comments. This research made
use of data products from the Midcourse Space Experiment.  Processing of
the data was funded by the Ballistic Missile Defense Organization with
additional support from NASA Office of Space Science.  This research has
made use of the NASA/IPAC Infrared Science Archive, which is
operated by the Jet Propulsion Laboratory, California Institute of
Technology, under contract with the National Aeronautics and Space
Administration, and the SIMBAD database, operated at CDS, Strasbourg, France
 GAIA is a Starlink derivative of the ESO Skycat catalogue
and image display tool.

\appendix 

\begin{figure*}
\epsfysize=23cm   \epsfbox[20 20 575 735]{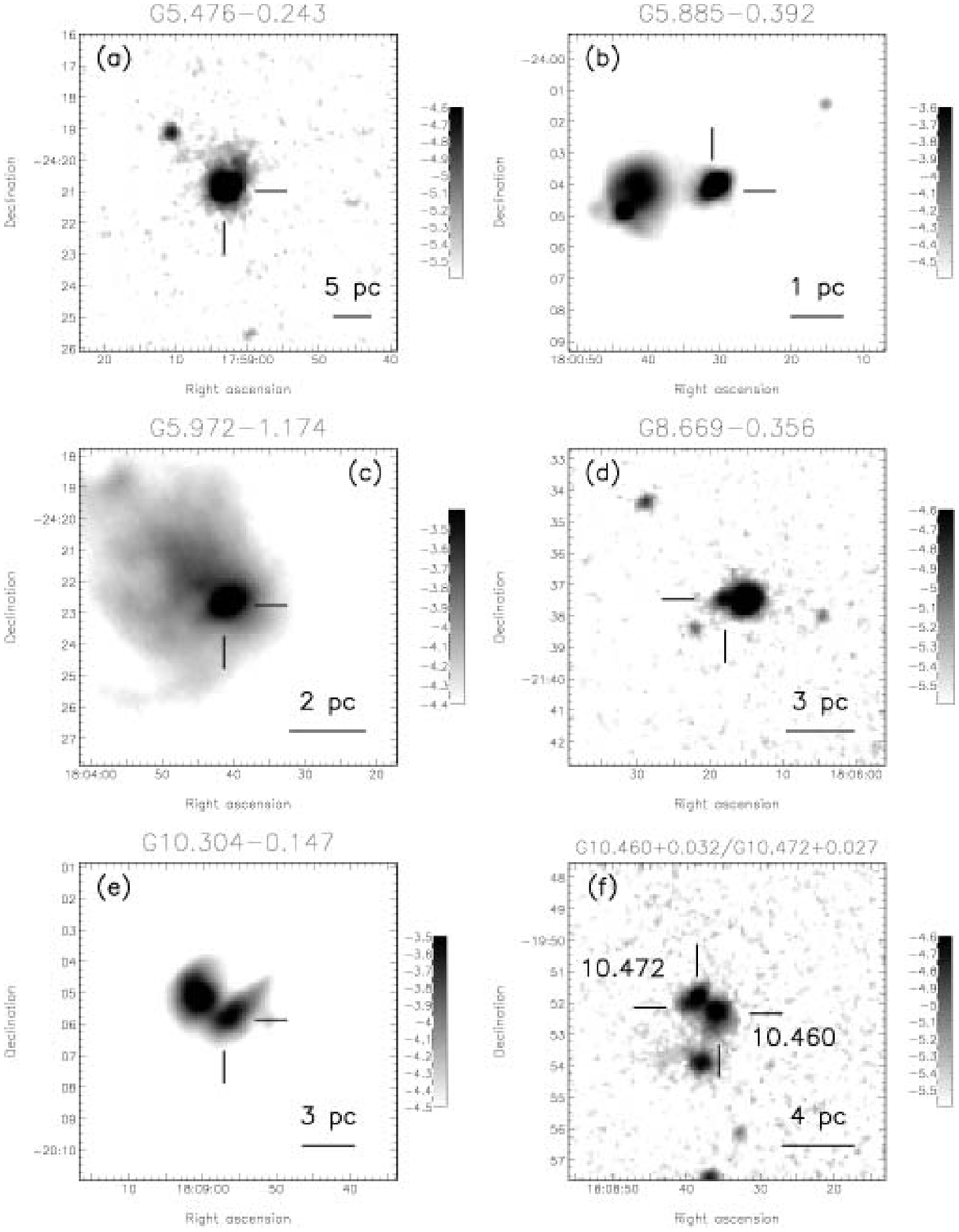}
{\bf Figure A1}. MSX Band E images of UCH\,{\sc ii} regions. Each field covers
a field-of-view of 10$\times$10 arcmin, and is presented in a logarithmic
intensity scale (units are W\,m$^{-2}$ sr$^{-1}$)
\label{fig1}
\end{figure*}

\begin{figure*}
\epsfysize=23cm   \epsfbox[20 20 575 735]{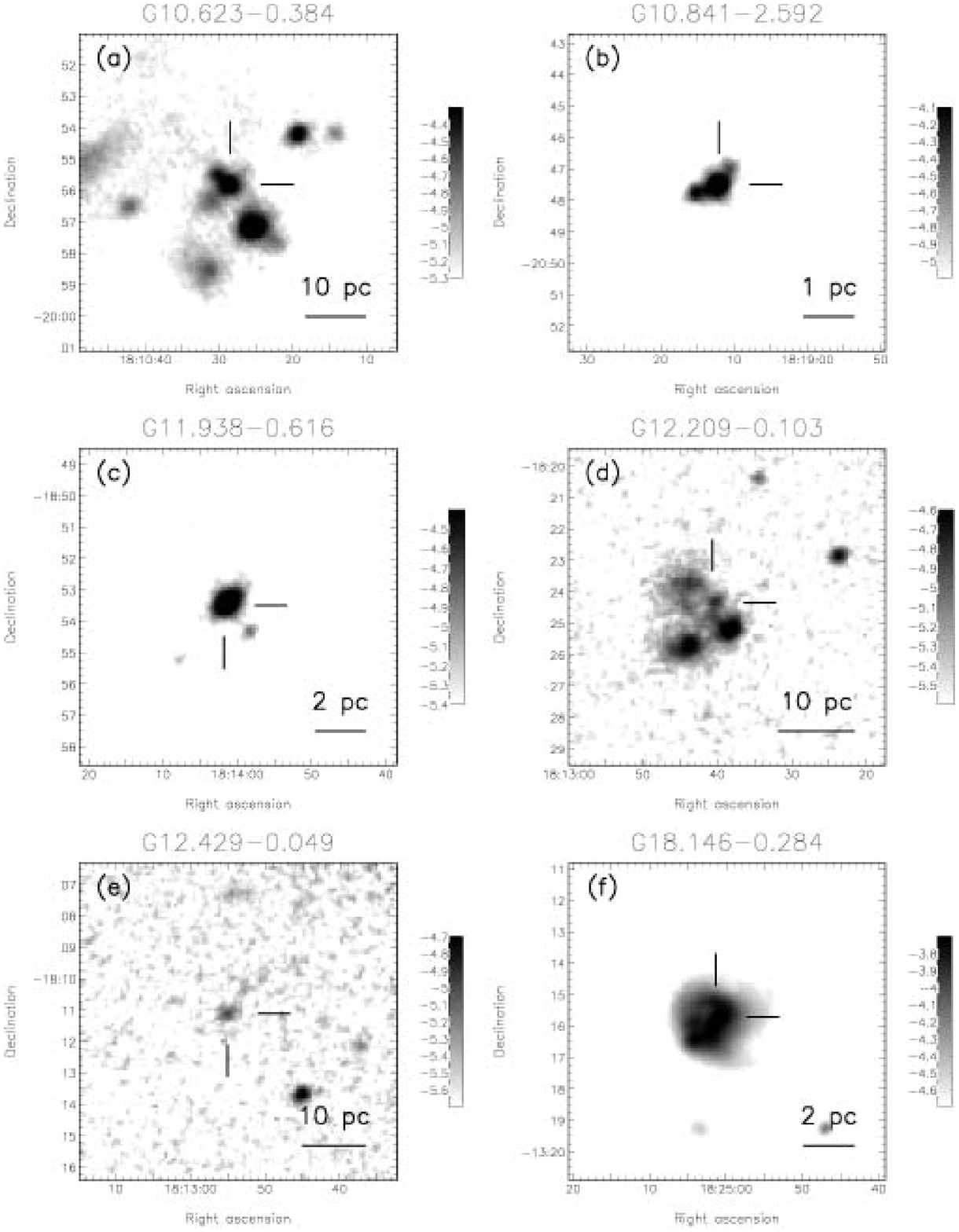}
{\bf Figure A2:} MSX Band E images of UCH\,{\sc ii} regions. Each field covers
a field-of-view of 10$\times$10 arcmin, and is presented in a logarithmic
intensity scale (units are W\,m$^{-2}$ sr$^{-1}$)
\label{fig2}
\end{figure*}

\begin{figure*}
\epsfysize=23cm   \epsfbox[20 20 575 735]{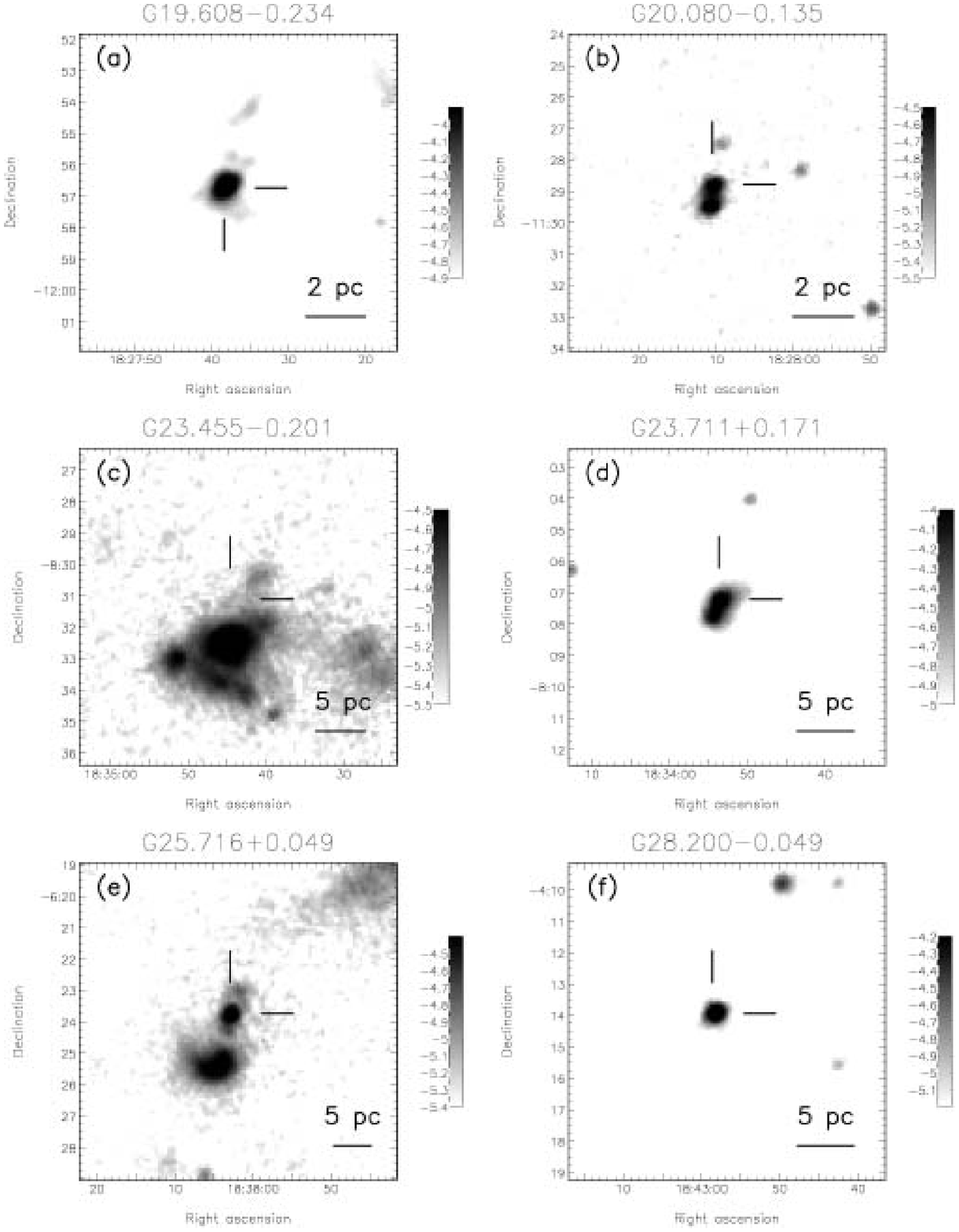}
{\bf Figure A3:} MSX Band E images of UCH\,{\sc ii} regions. Each field covers
a field-of-view of 10$\times$10 arcmin, and is presented in a logarithmic
intensity scale (units are W\,m$^{-2}$ sr$^{-1}$)
\label{fig3}
\end{figure*}

\begin{figure*}
\epsfysize=23cm   \epsfbox[20 20 575 735]{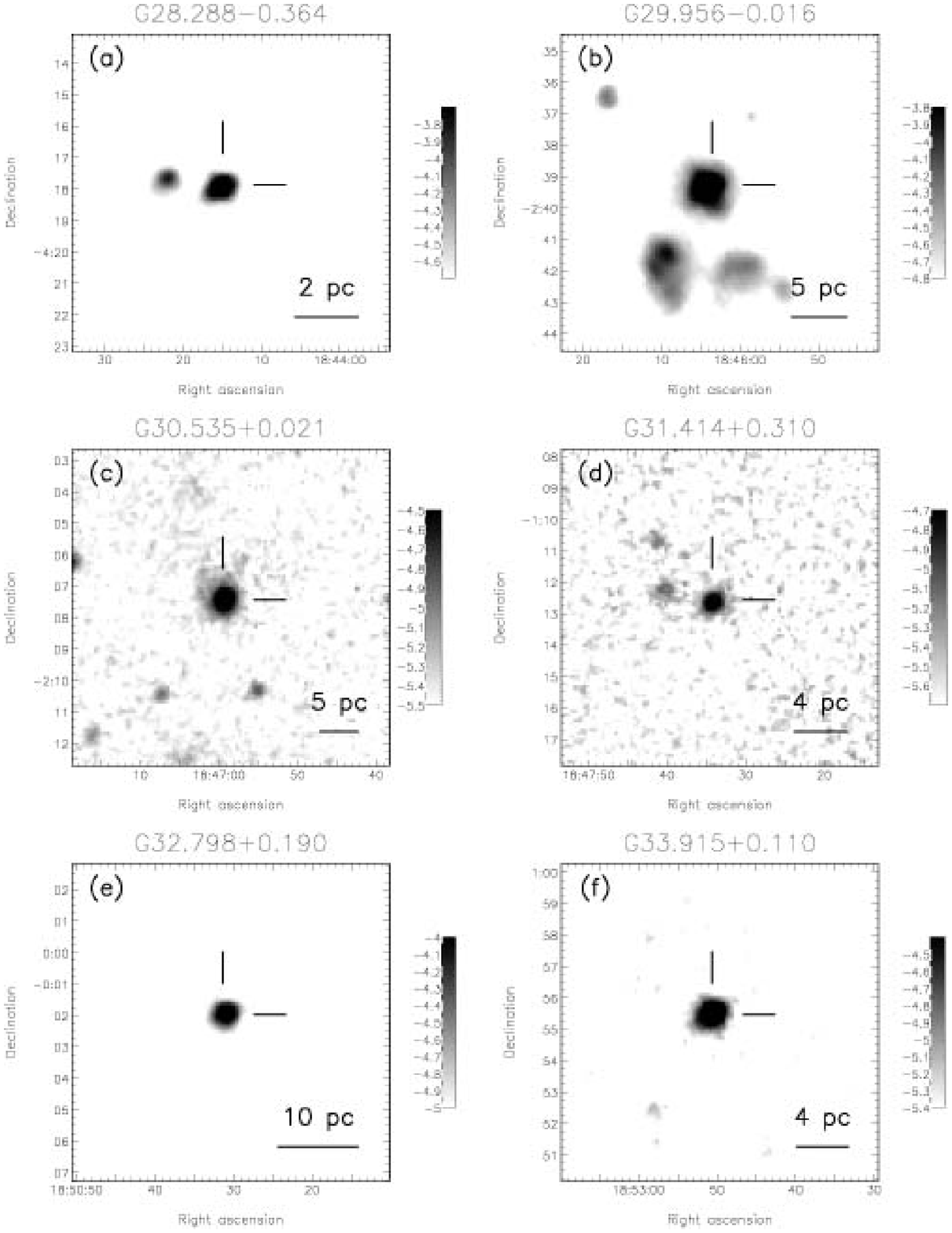}
{\bf Figure A4:} MSX Band E images of UCH\,{\sc ii} regions. Each field covers
a field-of-view of 10$\times$10 arcmin, and is presented in a logarithmic
intensity scale (units are W\,m$^{-2}$ sr$^{-1}$)
\label{fig4}
\end{figure*}

\begin{figure*}
\epsfysize=23cm   \epsfbox[20 20 575 735]{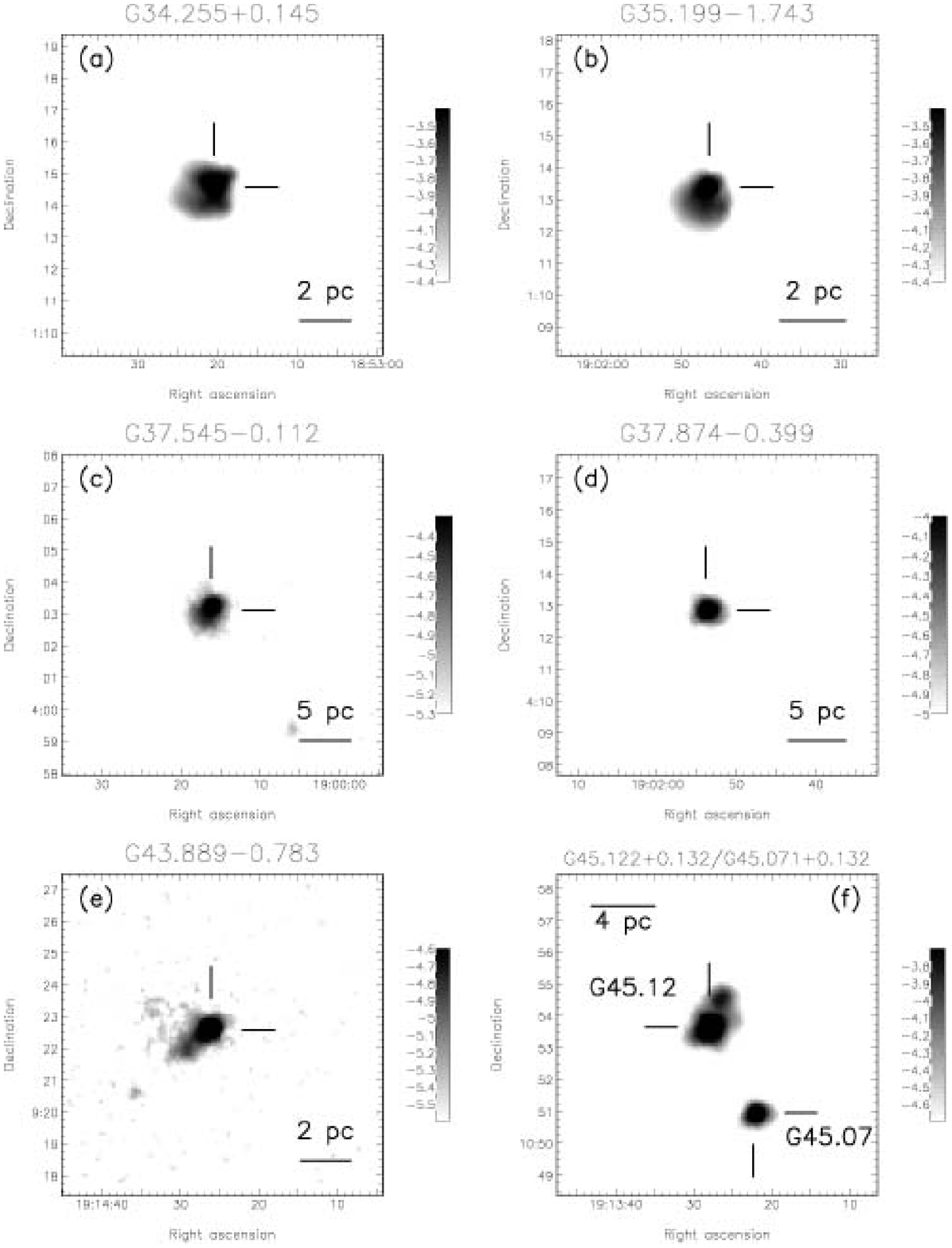}
{\bf Figure A5:}
MSX Band E images of UCH\,{\sc ii} regions. Each field covers
a field-of-view of 10$\times$10 arcmin, and is presented in a logarithmic
intensity scale (units are W\,m$^{-2}$ sr$^{-1}$)
\label{fig5}
\end{figure*}

\begin{figure*}
\epsfysize=23cm   \epsfbox[20 20 575 735]{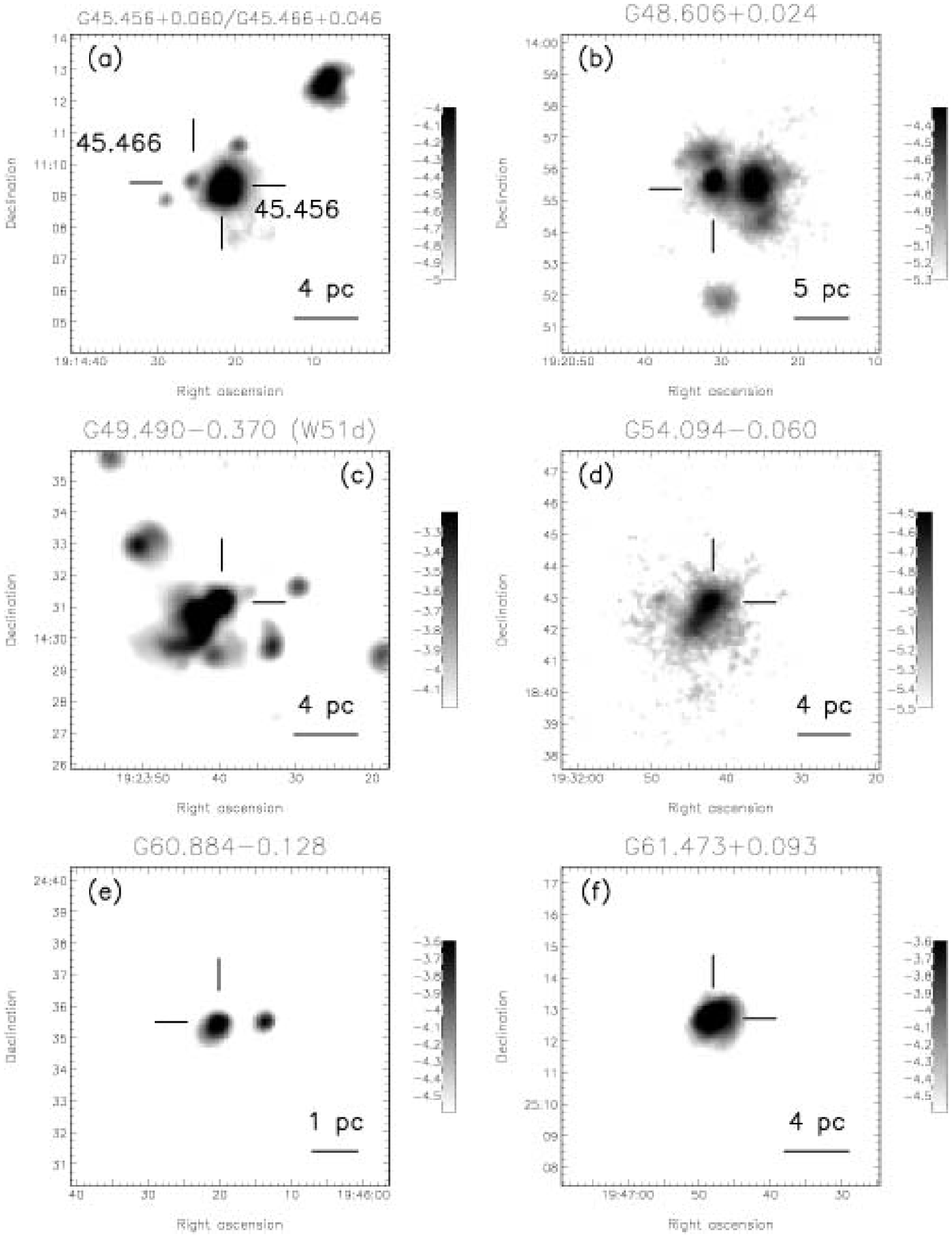}
{\bf Figure A6:}
MSX Band E images of UCH\,{\sc ii} regions. Each field covers
a field-of-view of 10$\times$10 arcmin, and is presented in a logarithmic
intensity scale (units are W\,m$^{-2}$ sr$^{-1}$)
\label{fig6}
\end{figure*}

\begin{figure*}
\epsfysize=23cm   \epsfbox[20 20 575 735]{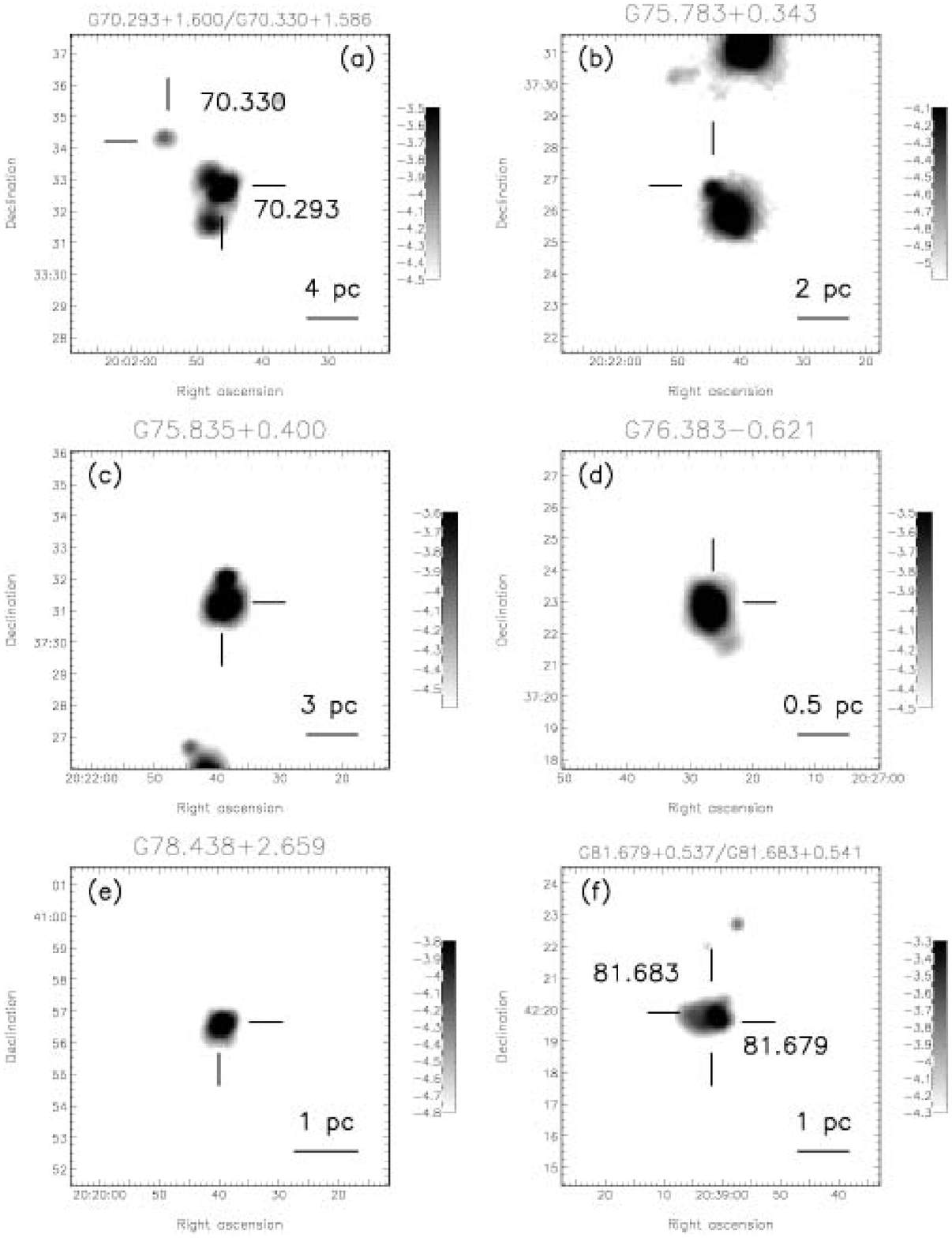}
{\bf Figure A7:}
MSX Band E images of UCH\,{\sc ii} regions. Each field covers
a field-of-view of 10$\times$10 arcmin, and is presented in a logarithmic
intensity scale (units are W\,m$^{-2}$ sr$^{-1}$)
\label{fig7}
\end{figure*}

\begin{figure*}
\epsfysize=23cm   \epsfbox[20 20 575 735]{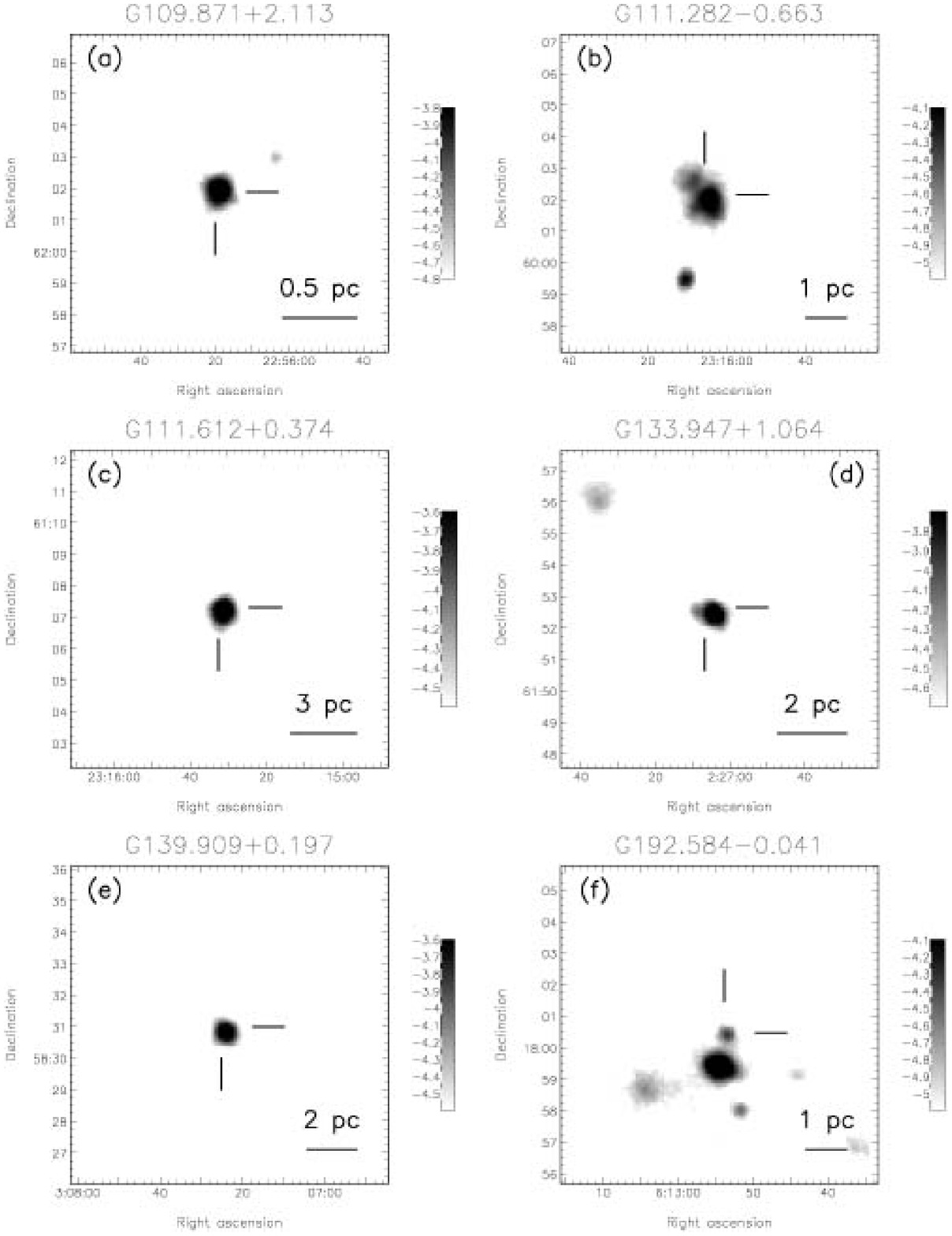}
{\bf Figure A8:} MSX Band E images of UCH\,{\sc ii} regions. Each field covers
a field-of-view of 10$\times$10 arcmin, and is presented in a logarithmic
intensity scale (units are W\,m$^{-2}$ sr$^{-1}$)
\label{fig8}
\end{figure*}

\bsp
\label{lastpage}

\begin{thebibliography}{99}

\bibitem{} Acord, J.M., Churchwell E., Wood D.O.S., 1998,ApJ 495, L107

\bibitem{} Araya E., Hofner P., Churchwell E., Kurtz S., 2002, ApJS 138, 63

\bibitem{} Churchwell E., 1999a, ARA\&A 40, 27

\bibitem{} Churchwell, E. 1999b, in The Origins of Stars and Planetary 
Systems, ed. C.J. Lada \& N.D. Kylafis (Dordrecht: Kluwer), p515

\bibitem{} Churchwell E., 2002, in Hot Star Workshop III: The Earliest
Stages of Massive Star Birth (ed. P.A. Crowther), ASP Conf. Ser 267, 3


\bibitem{} Churchwell, E., Walmsley, C.M., \& Cesaroni, R. 1990, A\&A Supp. 
83, 119

\bibitem{} Cohen, M., Walker R.G., Barlow, M.J., Deacon J.R., 1992, AJ, 
104, 1650

\bibitem{} Conti, P.S. \& Blum, R.D. 2002, in Hot Star Workshop III: The 
Earliest Stages of Massive Star Birth (ed. P.A. Crowther), ASP Conf. Ser 
267, 297

\bibitem{} Crowther P.A. (ed.), 2002, Hot Star Workshop III: The 
Earliest Stages of Massive Star Birth, ASP Conf. Ser 267

\bibitem{} De Buizer J.M., Watson A.M., Radomski J.T., Pina R.K., Telesco C.M.,
2002, ApJ 564, L101

\bibitem{} Draper, P.W., Gray N., \& Berry, D.S., 2001, 
Starlink User Note 214.9, Rutherford Appleton Laboratory, UK

\bibitem{} Egan M.P., et al. 1999, MSX Point Source Catalog Explanatory
Guide, AFRL-VS-TR-1999-1522, (Springfield: NTIS) 

\bibitem{} Feldt M., Stecklum, B., Henning Th., et al., 1998, A\&A 339, 759

\bibitem{} Feldt M., Stecklum, B., Henning Th., Launhardt R., Hayward T.L., 1999, A\&A 346, 243

\bibitem{} Fey A.L., Claussen M.J., Gaume R.A., Nedouuha G.E., Johnston K.J., 1992, AJ 103, 234

\bibitem{} Garay G., Rodriguez L.F., Moran J.F., Churchwell E., 1993, ApJ 418, 368

\bibitem{} Hatchell J., Fuller G.A., Millar T.J., Thompson M.A., Macdonald, G.H., 2000, A\&A 357, 637

\bibitem{} Hanson M.M., Luhman, K.L., Rieke G.H., 2002, ApJS 138, 35

\bibitem{} Hunter T.R., Phillips T.G., Menten, K.M., 1997, ApJ 478, 283

\bibitem{} Kahn F.D., 1974, A\&A 37, 149

\bibitem{} Kim K-T, Koo B-C, 2001, ApJ 549, 979

\bibitem{} Kraemer K.E., Jackson J.M., Deutsch L.K. et al. 2001, ApJ 561, 282

\bibitem{} Kurtz, S.E., Churchwell, E., \& Wood, D.O.S. 1994, ApJS, 91, 659

\bibitem{} Kurtz S.E., Watson A.M., Hofner P., Otte, B., 1999, ApJ 514, 232

\bibitem{} Kurtz, S.E., Cesaroni, R., Churchwell, E., Hofner, P., \& 
Walmsley, C.M. 2000 in Protostars and Planets IV, ed. V. Mannings, 
A.P.Boss,  ~\& S.S. Russell (Tucson: Univ. of Arizona Press), p299 


\bibitem{} Lumsden S.L., S.L., Hoare M.G., Oudmaijer R.D., Richards D., 2002, MNRAS 336, 621

\bibitem{} Lumsden, S.L., Puxley P.J., Hoare M.G., Moore T.J.T., Ridge, N.A., 2003, MNRAS in press
(astro-ph/0212135)

\bibitem{} Martins F., Schaerer D., Hillier D.J., 2002, A\&A 382, 999

\bibitem{} Morris P.M., van der Hucht K.A., Crowther P.A. et al. 2000, A\&A 353, 624

\bibitem{} Okamoto Y.K., Kataza H., Yamashita T. Miyata T, Onaka T., 2001, ApJ 553, 254

\bibitem{} Okamoto Y.K., Kataza H., Yamashita T. et al., 2003, ApJ, 584, 
368

\bibitem{} Osorio M., Lizano S., D'Alessio P., 1999, ApJ 525, 808

\bibitem{} Peeters E., Martin-Hernandez N.L., Damour F., Cox P., Roelfsema P.R. et
al. 2002, A\&A 381, 571

\bibitem{} Price S.D., Egan M.P., Carey S.J., Mizuno D.R., Kuchar T.A., 2001,
           AJ, 121, 2819

\bibitem{} Smith L.J., Norris R.P.F., Crowther P.A., 2002, MNRAS 337, 1309

\bibitem{} Stecklum B., Feldt, M. Richichi A. et al. 1997, ApJ 479, 339

\bibitem{} Stecklum B., Henning T., Feldt, M. et al. 1998, AJ 115, 767

\bibitem{} Stecklum B., Brandl B., Henning T. et al. 2002, A\&A 392, 1025


\bibitem{} Watson A.M., Hanson M.M., 1997, ApJ 490, L165

\bibitem{} Wolfire, M.G., Churchwell, E. 1994, ApJ, 427, 889

\bibitem{} Wood D.O.S., Churchwell E., 1989, ApJS, 69, 831

\end{thebibliography}
\end{document}